\documentclass[%
 reprint,
superscriptaddress,
 amsmath,amssymb,
 aps,
floatfix,
]{revtex4-2}

\usepackage{graphicx}
\usepackage{dcolumn}
\usepackage{bm}
\usepackage{xcolor}
\usepackage{hyperref}
\usepackage[normalem]{ulem}

\usepackage{placeins}   
\usepackage{balance}    

\usepackage{capt-of} 

\begin{document}







\title{Optimization of sequential therapies to maximize
extinction of resistant bacteria through collateral
sensitivity}

\author{Javier Molina-Hernández}
\affiliation{Universidad Carlos III de Madrid, Departamento de Matemáticas, Leganés, Spain}
\affiliation{Grupo Interdisciplinar de Sistemas Complejos (GISC)}
\author{José A. Cuesta}
\affiliation{Universidad Carlos III de Madrid, Departamento de Matemáticas, Leganés, Spain}
\affiliation{Grupo Interdisciplinar de Sistemas Complejos (GISC)}
\affiliation{Instituto de Biocomputación y Física de Sistemas Complejos, Universidad de Zaragoza, Zaragoza, Spain}
\author{Beatriz Pascual-Escudero}
\affiliation{Universidad Politécnica de Madrid}
\author{Saúl Ares}
\email{saul.ares@csic.es}
\affiliation{Centro Nacional de Biotecnologia (CNB), CSIC, Madrid, Spain}
\affiliation{Grupo Interdisciplinar de Sistemas Complejos (GISC)}
\author{Pablo Catalán}
\email{pcatalan@math.uc3m.es}
\affiliation{Universidad Carlos III de Madrid, Departamento de Matemáticas, Leganés, Spain}
\affiliation{Grupo Interdisciplinar de Sistemas Complejos (GISC)}


\begin{abstract}
Antimicrobial resistance (AMR) threatens global health. A promising and underexplored strategy to tackle this problem is sequential therapies exploiting collateral sensitivity (CS), whereby resistance to one drug increases sensitivity to another. Here, we develop a four-genotype stochastic birth–death model with two bacteriostatic antibiotics to identify switching periods that maximize bacterial extinction under subinhibitory concentrations. We show that extinction probability depends nonlinearly on switching period, with stepwise increases aligned to discrete switch events: fast sequential therapies are suboptimal as they do not allow for the evolution of resistance, a key ingredient in these therapies. A geometric distribution framework accurately predicts cumulative extinction probabilities, where the per-switch extinction probability rises with switching period. We further derive a heuristic approximation for the extinction probability based on times to fixation of single-resistant mutants. Sensitivity analyses reveal that strong reciprocal CS is required for this strategy to work, and we explore how increasing antibiotic doses and higher mutation rates modulate extinction in a nonmonotonic manner. Finally, we discuss how longer therapies maximize extinction but also cause higher resistance, leading to a Pareto front of optimal switching periods. Our results provide quantitative design principles for \emph{in vitro} and clinical sequential antibiotic therapies, underscoring the potential of CS‐guided regimens to suppress resistance evolution and eradicate infections.
\end{abstract}

\date{\today}

\maketitle





\section{Introduction}
\label{sec:intro}

Antimicrobial resistance (AMR) is rising rapidly \cite{catalan2022seeking}, leading to higher rates of uncontrolled infections that contribute significantly to both patient mortality and healthcare costs \cite{naghavi2024global}. The development of new antibiotics is not fast enough to counteract this problem \cite{miethke2021towards} and, although new technologies can help accelerate discovery \cite{liu2023deep}, this is not guaranteed to solve the problem, as resistance to new antibiotics evolves soon after or even before their deployment in the clinic \cite{martins2025antibiotic, daruka2025eskape}.

An alternative strategy to combat AMR is to develop multidrug treatments \cite{baym2016multidrug}, unlocking access to large combinatorial treatment spaces. Combination therapies are the most explored alternative, where two or more antibiotics are deployed simultaneously \cite{tyers2019drug}. These therapies can prevent the rise of AMR \cite{roemhild2022physiology}, especially if the antibiotics are chosen \textit{ad hoc} for a particular pathogen \cite{rosenkilde2019collateral} or based on their interaction profiles \cite{kavvcivc2020mechanisms}. However, combination therapies are not without disadvantages: the total concentration of antibiotics introduced in the patient is higher than monotherapies, and so there is the risk of toxic effects \cite{tamma2012combination}. Moreover, combinations have sometimes been found to accelerate, rather than slow down, the appearance of resistance \cite{pena2013most, vestergaard2016antibiotic}.

A less explored alternative to combination therapies is sequential therapies, in which several antibiotics are administered one after the other instead of simultaneously \cite{roemhild2019evolutionary}. This strategy is based on the phenomenon of collateral sensitivity (CS), in which bacterial resistance to one antibiotic increases its sensitivity to another. CS has been found in many bacterial species and antibiotic classes \cite{barbosa2017alternative, imamovic2018drug, podnecky2018conserved, hernando2020rapid}, suggesting a promising avenue to develop treatments that can eradicate pathogenic bacterial populations \cite{fuentes2015using, batra2021high}. The sequential framework opens the door to mathematical optimization approaches, where we seek the optimal sequence that maximizes the eradication of the population or minimizes the evolution of resistance \cite{beardmore2017antibiotic, maltas2019pervasive, morsky2022suppressing, katriel2024optimizing, weaver2024reinforcement, maltas2025dynamic}.


Here, we seek antibiotic switching protocols that maximize bacterial extinction in a stochastic population model where only four genotypes and two antibiotics are considered, similarly to previous work \cite{aulin2021design}. While transmission dynamics between different patients are a critical element driving the AMR crisis and have been studied in depth before \cite{bonhoeffer1997evaluating, angst2021comparing, muetter2024impact}, we focus on within-host evolutionary dynamics, a key yet poorly understood aspect of AMR \cite{shepherd2024ecological} where \textit{de novo} mutations can appear in the span of days or weeks \cite{mwangi2007tracking, blair2015acrb, wheatley2021rapid, chung2022rapid}. 

We show that sequential therapies based on strong CS can lead to bacterial eradication even at subinhibitory concentrations, provided that the correct switching period is used. The dependence of bacterial extinction on the switching period is explained by population composition, paving the way for optimization based on population metrics. Our results are contingent on the existence of strong CS, and we observe nonmonotonic relationships between extinction rates and antibiotic dose, on one hand, and mutation rates, on the other, which we can explain using our model. We end by searching for optimal switching periods that both maximize extinction and minimize the appearance of resistance, finding there is a tradeoff between both objectives, leading to a Pareto front of optimal therapies. Our work suggests that sequential therapies are a rich opportunity to explore potentially successful strategies that will help tackle the antibiotic crisis.



















\section{The model}

\begin{figure}[t]
  \centering
    \includegraphics[width=0.45\textwidth]{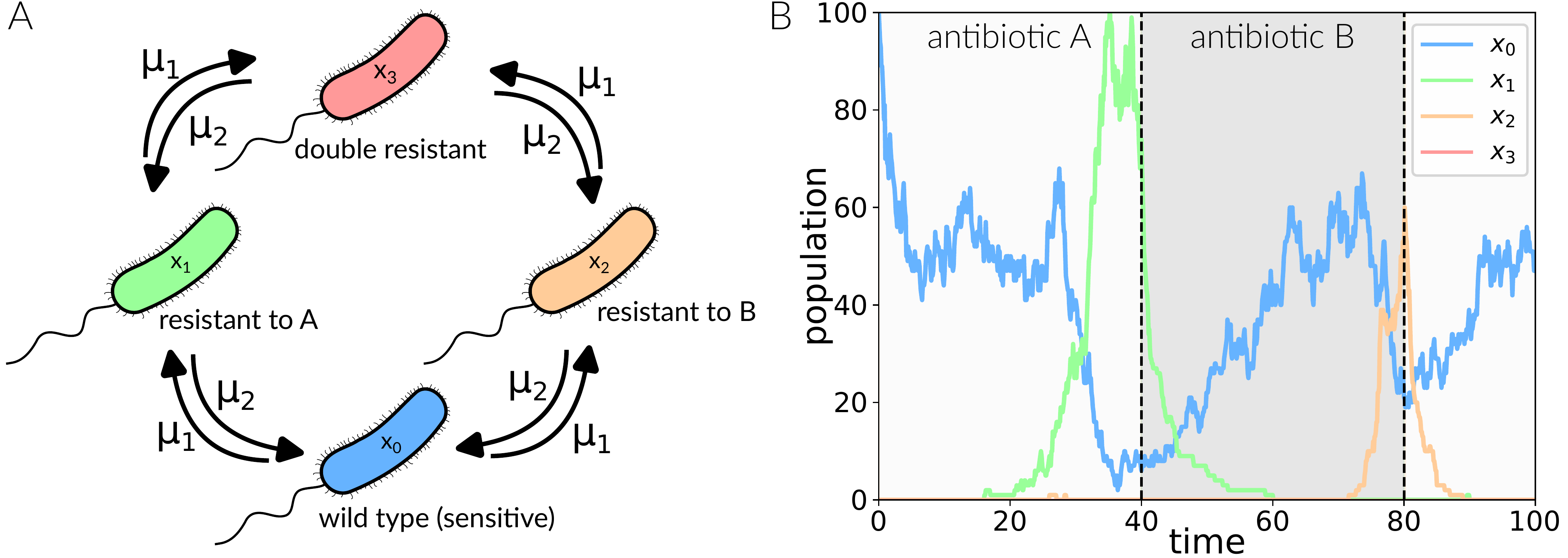}
    \caption{Four-genotype model. (a) We consider four genotypes:
    $x_0$ (blue), susceptible to both antibiotics, $x_1$ (green), resistant to antibiotic $A$ but
    susceptible to $B$, $x_2$ (orange) resistant to $B$ and susceptible to $A$,
    and $x_3$ (red), resistant to both antibiotics. Mutation rates between the genotypes
    are indicated next to the corresponding arrows. (b) Illustrative trajectory. We start our simulations with antibiotic $A$ and, after some time $\tau$, we switch to antibiotic $B$, and repeat the process.}
    \label{fig:network}
\end{figure}

We will study bacterial population dynamics within a single patient under two bacteriostatic
antibiotics $A$ and $B$, using a four-genotype birth-death stochastic
model. The four types are: $x_0$, susceptible to both antibiotics;
$x_1$, resistant to $A$ but susceptible to $B$; $x_2$, resistant to
$B$ but susceptible to $A$; and $x_3$, resistant to both (Fig.~\ref{fig:network}(a)). We consider stochastic transitions
(mutations) between types, with rates $\mu_1$ for the acquisition of
resistance and $\mu_2$ for the loss of resistance \cite{GilGil2024SCV_HR}. Mutations from $x_0$ to
$x_3$ are not permitted, although their introduction does not qualitatively change our results. We do not consider transmission of new strains from other patients, focusing instead on within-host evolutionary dynamics. In what follows, we will refer to the population of genotype $x_i$ as $N_i$.


Birth rates are different for each type, and depend on the antibiotic
we are using. Type $x_0$ reproduces with rate $\beta_{0,A}=k_A \beta$ under
antibiotic $A$ and with rate $\beta_{0,B}=k_B \beta$ under antibiotic $B$, where $k_A, k_B \in [0,1]$ is a measure of antibiotic inhibition: the antibiotic effect is stronger the lower the value of $k$. In
what follows, we make the simplifying assumption $k_A=k_B=k$ and leave the analysis of
other scenarios for future works. Type $x_1$ reproduces with birth rate
$\beta_{1,A}=\beta$ under antibiotic $A$ and with rate $\beta_{1,B}=k_{\rm  CS} k_B \beta$ under
antibiotic $B$, where $k_{\rm  CS} \in [0,1]$ is a measure of the lack of collateral
sensitivity: when $k_{\rm  CS} \to 1$ CS is absent,
whereas when $k_{\rm  CS} \to 0$ it is very strong. Birth rates for type $x_2$
are symmetrical to those of type $x_1$. Birth rates for type $x_3$ are $\beta_{3,A}=\beta_{3,B}=\beta$ in
both antibiotics. For simplicity of notation, we will
nondimensionalize time by defining the variable $\beta t$,
which is equivalent to setting $\beta=1$.

Death rates are equal for all types and antibiotics and equal
to $\delta_{i,A}=\delta_{i,B}=\gamma N, i=0,1,2,3$, where $N=N_0+N_1+N_2+N_3$ is the total
population, simulating limited resources. Note that $\gamma\beta_{i,j}^{-1}$ is the inverse of the carrying
capacity in logistic models. We fix $\gamma=0.01$ (one death per
one hundred births). 


For computational efficiency, in order to simulate the stochastic model we will use the tau-leaping
algorithm \cite{gillespie2007stochastic}, which approximates the dynamics of the birth-death process by taking small time increments $dt$ and generating
pseudo-random Poisson-distributed numbers for all reactions: births,
deaths, and mutations. Population sizes are updated accordingly, and the
process is repeated until desired.  Note that using Gillespie's algorithm \cite{gillespie1976general}, which simulates the model exactly, does not qualitatively change our results (see Appendix~\ref{si:gillespie}, Fig.~\ref{fig:supp-gillespie}).


We start our simulations with
antibiotic $A$ and initial population $N_0=k\cdot\gamma^{-1}, N_1=N_2=N_3=0$, i.e., initially there is no resistance and the population of susceptible is at the carrying capacity. After some elapsed time $\tau$ (switching period) we switch from antibiotic $A$ to $B$,
and continue the process. At time $2\tau$ we switch back to $A$ and so
on (Fig.~\ref{fig:network}(b)). We are
interested in studying how this parameter $\tau$, the period of
antibiotic switching, affects the probability that the population
becomes extinct at the end of treatment. 

We study treatments of different $\tau$ ranging from zero to $100$ time units, with a fixed treatment duration $T=100$. In our framework, treatment duration $T$ should be understood as corresponding to a typical antibiotic regime in a clinical context: for instance, a person taking one pill every $h$ hours for a total of $T$ hours, with the final dose administered at time $T$. However, the pharmacodynamic effects of the antibiotic are not expected to vanish instantaneously at $T$, since this marks the time of the last administered dose rather than the cessation of its biological activity. To account for this, we simulate an additional period under the final antibiotic (i.e., with no further switches) beyond time $T$, and evaluate extinction probabilities based on this extended simulation. This is particularly relevant for values of $\tau$ close to $T$, where the last antibiotic may have been recently applied and its impact is still unfolding.

For simulation purposes, we mark a population as
extinct whenever $N < 0.05\gamma^{-1}$~\cite{Ankomah2014Collaboration}, and our results remain qualitatively unchanged by reasonable changes in this threshold (Appendix~\ref{si:gillespie} and Appendix~\ref{si:gillespie}, Fig.~\ref{fig:supp-thresholds}).
%

\begin{figure*}[t]
  \centerline{\includegraphics[angle=0, width=1\textwidth]{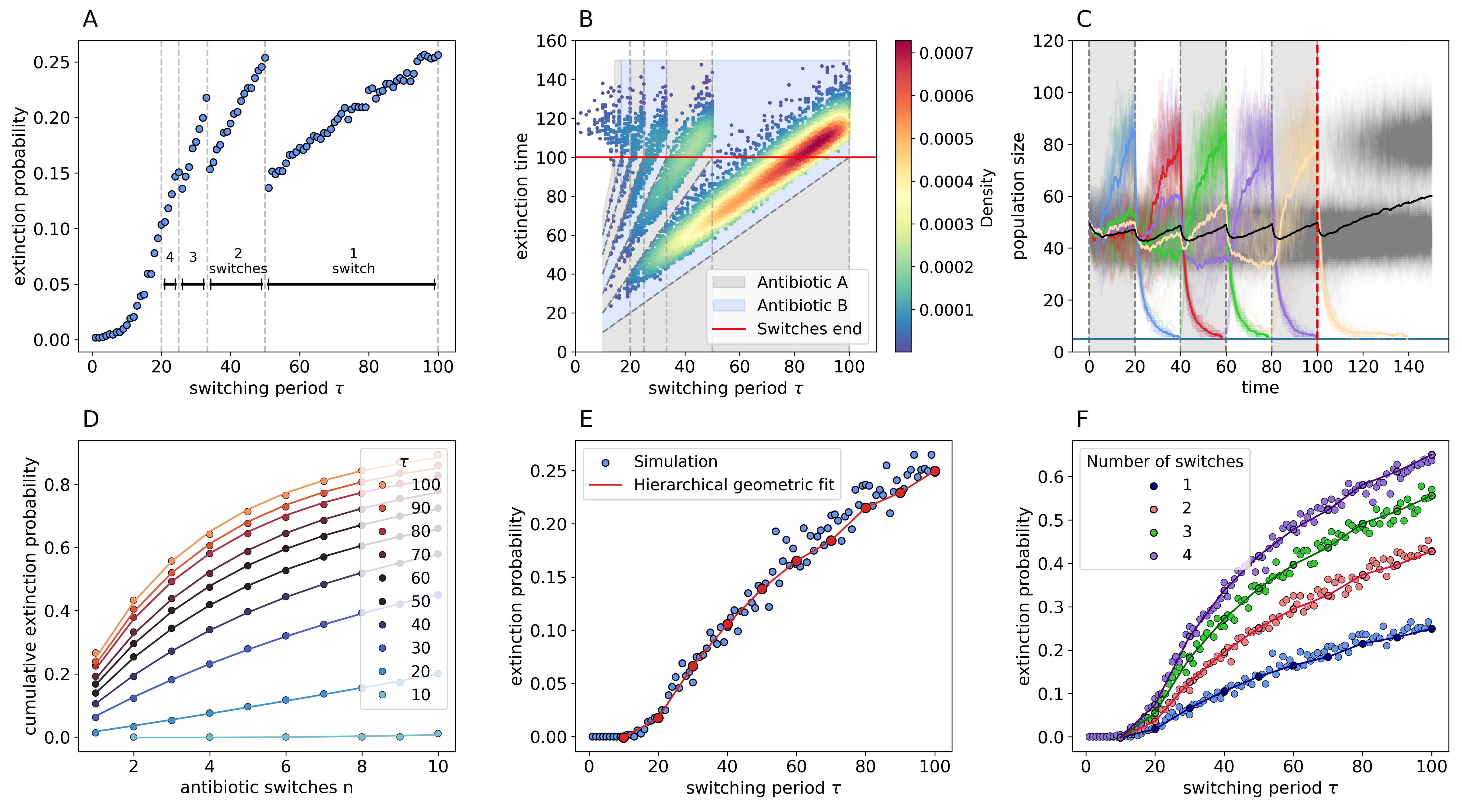}}
\caption[]{Sequential therapies with subinhibitory antibiotic concentrations cause extinction for a wide range of switching periods.
(a) Probability of extinction at the end of the treatment as switching periods vary. 
The intervals between two vertical lines share the same number of treatment cycles. $10,000$ trajectories were used to estimate the probability as a function of $\tau$.
(b) Distribution of extinction times as a function of switching periods. Each point corresponds to the time a simulation became extinct. The colors represent the density of these events. The background color represents the antibiotic used. The red line represents the end of the time we allow for switching antibiotics.
(c) One thousand individual trajectories switching treatment every 20 time units. Blue, red, green, purple, and orange represent the mean of trajectories that go extinct upon switching antibiotics at different switching events. Black represents the mean of those trajectories that do not become extinct. Dashed lines indicate the antibiotic switch. 
(d) Cumulative extinction probability over 10 treatment switches with different switching periods $\tau$. Points represent extinction probabilities estimated through simulation, while the solid line reflects a fit to a hierarchical geometric distribution.
(e) Extinction probability for populations undergoing one antibiotic switch; blue points represent the simulation, and the red points are the fitted extinction probabilities for the corresponding $\tau$ using the hierarchical geometric model (the red line is a guide to the eye).
(f) Predicted extinction probabilities for populations undergoing one to four antibiotic switches using the hierarchical geometric model. Simulations were performed with final times $\tau \cdot $(number of switches)$+ 50$. Parameter values in Appendix~\ref{app:table1}, Table~\ref{tab:parameters}.}
\label{fig:panel2}
\end{figure*}

\section{Results}

\subsection{Sequential treatments with strong collateral sensitivity result in bacterial eradication even with subinhibitory antibiotic concentrations}

We start our study with subinhibitory antibiotic concentrations: $k_A=k_B=0.5$, usually called the half-maximum inhibitory concentration or IC50, and consider strong CS, $k_{\rm  CS}=0.05$. This may seem like a strange place to begin, as these subinhibitory doses are usually thought to promote the evolution of resistance \cite{andersson2012evolution,reding2021antibiotic}. However, our simulations produce a wide range of switching periods $\tau$ that result in frequent eradication of the bacterial population. Fig.~\ref{fig:panel2}(a) shows that the probability that a population becomes extinct at the end of the treatment depends non-trivially on $\tau$. First, we observe that there are no extinctions when $\tau \to 0$ (Fig.~\ref{fig:panel2}(a,b)). As $\tau$ increases, there is a rapid increase in the probability of extinction, with two maxima at $\tau=50$ and $\tau=100$. The extinction probability shows some sharp discontinuities, which are due to a change in the number of antibiotic switches: extinction probabilities increase discontinuously when a new antibiotic switch is introduced. For instance, when $\tau$ is decreasing from $100$, the extinction probability suddenly increases when we reach $\tau=50$, where the number of antibiotic switches increases from 1 to 2. Similarly, when we cross the threshold $\tau=33.3$ there is another increase in extinction probabilities, associated with an increase in the number of switches from 2 to 3. Conversely, in the absence of switches (i.e. $\tau=0$ or $\tau > T$) there are no extinctions, which is to be expected since we are studying IC50 concentrations. Moreover, when studying extinction times, we observe that, for a fixed switching period $\tau$, extinction events occur more frequently after a short transient period following the switching time (Fig.~\ref{fig:panel2}(b)) and never before the first switch. In order to better illustrate this phenomenon, we examine a collection of trajectories for fixed $\tau=20$ (Fig.~\ref{fig:panel2}(c)): some trajectories become extinct after the first switch (marked in blue in Fig.~\ref{fig:panel2}(c)), some after the second (red), third (green), fourth (purple), and fifth (orange) switches. Populations that do not become extinct (black) eventually gain resistance.

\begin{figure*}[t] 
  \centerline{\includegraphics[angle=0, width=1\textwidth]{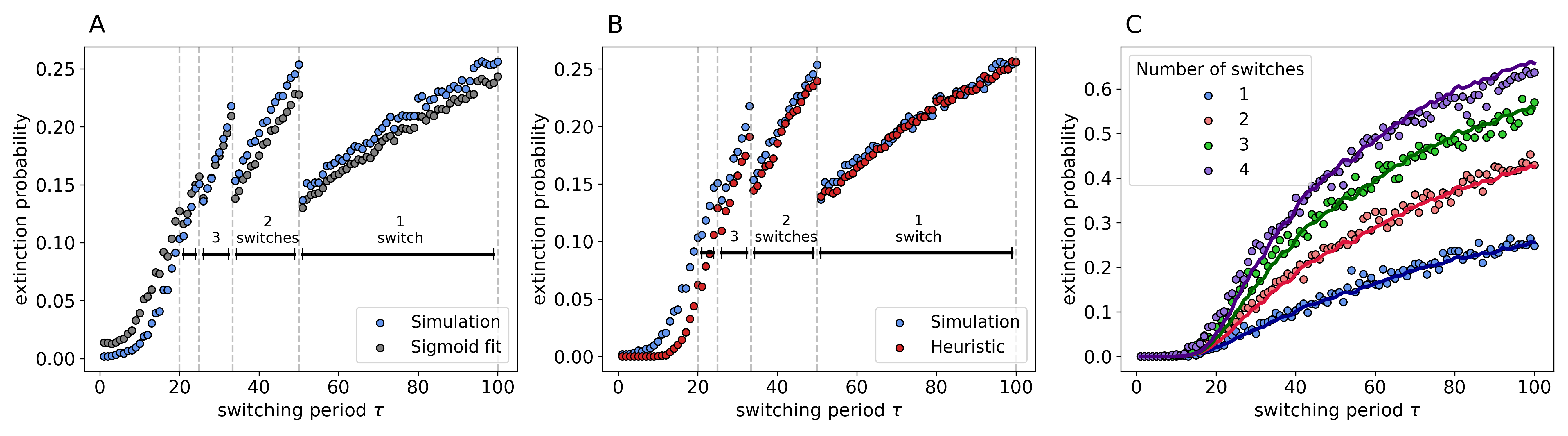}}
\caption[]{Heuristic approximation for the extinction probability. 
(a) Sigmoid fit for the extinction probability at the end of the treatment, using the population composition before the switch. Blue circles represent the simulation; gray circles are the prediction of the sigmoid function fitted with the population before the switches.
(b) Heuristic fit for the same probability as in (a) using fixed final time treatments, 
and considering several switching periods $\tau$.
(c) Heuristic fit for the extinction probability under different number of antibiotic switches. Circles are the values of extinction probabilities measured in simulations; each color represents a different number of antibiotic changes, as indicated in the legend. The solid lines indicate the estimated value of the probability of extinction for each $\tau$, using the heuristic to estimate $p$. Parameter values in Appendix~\ref{app:table1}, Table~\ref{tab:parameters}.}
\label{fig:heuristic}
\end{figure*}

Since extinction events occur shortly after switching antibiotics, we can visualize the extinction process as a coin-tossing game where the probability of getting heads is $p$: each time the antibiotic switches, we toss a coin. If the result is heads, the population becomes extinct. The probability that the extinction event occurs exactly after $n$ switches would then be given by the geometric probability distribution $(1-p)^{n-1} p$, where $p$ is the probability that the population becomes extinct after one antibiotic switch, and which may be dependent on $\tau$. The cumulative extinction probability, i.e. the probability that the population has become extinct by the $n$th switch, is given by $1-(1-p)^{n}$. The main hypothesis of this model is that $p$ is constant throughout the process, which will be true if the time between switching is sufficiently large so that the population reaches some kind of stationary state that is independent of the switching event. 

Fitting the cumulative extinction probability obtained from simulations for different $\tau$ to equation $1-(1-p)^{n}$ results in good agreement (Appendix~\ref{sec:hierarchical}, Fig.~\ref{fig:supp-noneq}(a)). However, the fits are not perfect: they tend to underestimate the extinction probabilities when the number of switches is small, and vice versa. This suggests that $p$ is not constant and slightly depends on the number of switches. We capture this behavior with a hierarchical geometric model where $p=\alpha \tau + \beta n$ can change with both switching period $\tau$ and the number of switches $n$, resulting in a great fit to the data (Fig.~\ref{fig:panel2}(d), Appendix~\ref{sec:hierarchical}). The values of $p$ obtained with our fits are shown in Fig.~\ref{fig:panel2}(e) (red line) as a function of $\tau$ and compared with the simulated extinction probabilities for treatments undergoing only one switch (blue circles): the probability of extinction per switching event is close to zero for $\tau < 20$, consistent with what we observed in Fig.~\ref{fig:panel2}(a), and increases as $\tau$ grows. In other words, the longer we wait until we switch antibiotics, the higher the probability that the population becomes extinct. The agreement between the simulated per-switch extinction (blue circles) and the hierarchical geometric fit (red lines) is quite good, given the simplicity of the model. We can use this fit to predict extinction times after two or more switches, by calculating the corresponding probability using the geometric distribution, with very good agreement (Fig.~\ref{fig:panel2}(f)), supporting our hypothesis.

\subsection{Heuristic explanation for the change in extinction probabilities}

We turn now to give an explanation for the observed extinction patterns by finding out which variables explain the dependence of the extinction probability on the switching period $\tau$. Fig.~\ref{fig:panel2}(c) hints that, in those populations that become extinct, right before the switch the population had become dominated by the single-resistant populations ($x_1$ or $x_2$) reaching a carrying capacity close to $100$. In contrast, the populations where extinction does not happen (marked in gray in Fig.~\ref{fig:panel2}(c)) are dominated by $x_0$, whose carrying capacity is $50$. Our hypothesis is that, for small $\tau$, resistant mutants have hardly any time to appear in the population before the antibiotic is switched and, due to the strong CS, they are rapidly invaded after the switch by type $x_0$, which does not go extinct under IC50 concentrations. As $\tau$ increases, however, the likelihood that either $x_1$ or $x_2$ rise in the population grows, and therefore, when the antibiotic switches, there is a chance that the whole population goes extinct.

\begin{figure*}[t] 
  \centerline{\includegraphics[angle=0, width=1.0\textwidth]{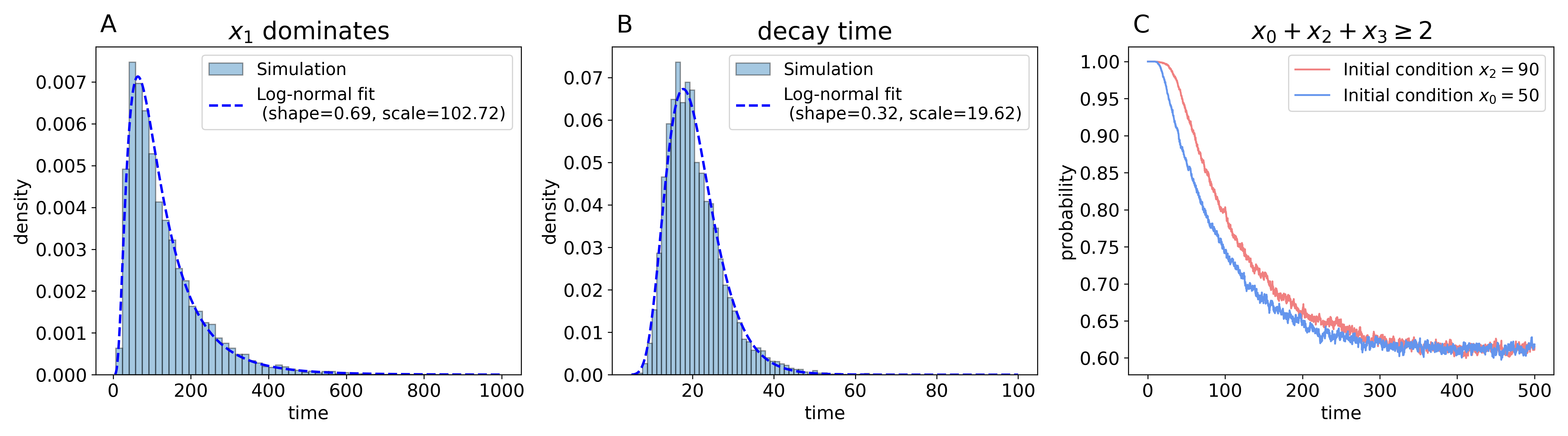}}
\caption[]{Distributions for key transition times in the model. 
Simulated distributions (blue histograms) and fitted lognormal probability density functions (dashed curves) for two temporal processes underlying the heuristic extinction estimate. 
(a) Time until the system transitions between stable states following an antibiotic change. Starting from $x_2(0)=80$ and $x_i(0)=0$ for $i=0,1,3$ under antibiotic $A$ until $x_1(t)>80$. 
(b) Time to extinction under a new antibiotic, measured only for simulations where extinction occurs. 
(c) Time dependent probability of having $x_0+x_2+x_3 \geq 2$ starting from the initial conditions indicated by colors in the legend.
$10000$ trajectories were simulated for estimating these distributions.}
\label{fig:gammafits}
\end{figure*}

This discussion suggests that we should be able to predict extinction probabilities from the composition of the population before the switch. We have fitted the probability $p$ that a population becomes extinct after an antibiotic switch to a sigmoid function $p=\big(1+e^{\mathbf W^T \mathbf N}\big)^{-1}$ where $\mathbf W$ is a parameter vector and $\mathbf N=(N_0, \dots, N_3)$ contains the population abundances right before the antibiotic switch (Fig.~\ref{fig:heuristic}(a), Appendix~\ref{sec:sigmoid}). The fitted parameters confirm our earlier intuition (Appendix~\ref{sec:sigmoid}, Table~\ref{tab:parameterssigmoid1}): $p$ decreases when any genotype other than the single resistant increases. For example, if the population before the switch is $N_0=2, N_1=95, N_3=0$, then $p \approx 0.55$, but introducing one $x_3$ individual yields $p \approx 0.26$. This indicates that the extinction probability is largely determined by whether the population contains cells other than the currently dominant resistant genotype---the presence of even a small fraction of any other genotypes substantially decreases the chances of extinction.


However, it is unrealistic to assume that complete compositional data will be available in clinical or \textit{in vitro} settings to decide when to switch antibiotics. We therefore propose a heuristic approach to approximate $p$ using three key time distributions obtained from the dynamics of our four-genotype system. Specifically, we measured: (1) the distribution of times for single-resistant mutants to dominate the population (defined as exceeding $80\%$) after an antibiotic switch; (2) the distribution of extinction times following an antibiotic switch; and (3) the probability that the population contains cells other than the single-resistant mutant for the antibiotic currently in use, a choice motivated by the sigmoid fit, as previously discussed. 

The first two distributions were well-approximated by lognormal distributions (Fig.~\ref{fig:gammafits}), and all three can in principle be measured \textit{in vitro} to obtain a practical estimate of optimal switching strategies. Building on these distributions and the previous insights gained from our geometric model, we derived an analytical expression for the extinction probability $p$ as a function of $\tau$:
\begin{equation}
    p(\tau) = p_d(\tau)\bigl[1 - p_r(\tau)\bigr],
\label{eq:heuristica}
\end{equation}
where $p_d(\tau)$ is the probability that a decaying population becomes extinct within time $\tau$ (Fig.~\ref{fig:gammafits}(b)), and $p_r(\tau)$ is the probability that a single-resistant mutant is not completely dominant in the population, given by
\begin{align*}
    p_r(\tau) =\; p_{\mathrm{x_1}}(\tau)\,&\cdot \Pr(N_0+N_2+N_3 \geq 2 \mid \tau, N_2(0)=90) \\
    + \bigl[1-p_{\mathrm{x_1}}(\tau)\bigr]&\cdot \Pr(N_0+N_2+N_3 \geq 2 \mid \tau, N_0(0)=50)
\end{align*}
with $p_{\mathrm{x_1}}(\tau)$ denoting the probability that genotype $x_1$ dominates the population at time $\tau$, given that the system started with $x_2$ in dominance (Fig.~\ref{fig:gammafits}(a)). Intuitively, for short times the initial condition $N_0(0)=50$ provides a good approximation, while for longer times the condition $N_2(0)=90$ becomes more accurate. The weighting between these two scenarios is naturally captured by the distribution of takeover times of the single-resistant strain, represented by $p_{\mathrm{x_1}}(\tau)$. 

\begin{figure*}[!t] 
  \centerline{\includegraphics[angle=0, width=1.0\textwidth]{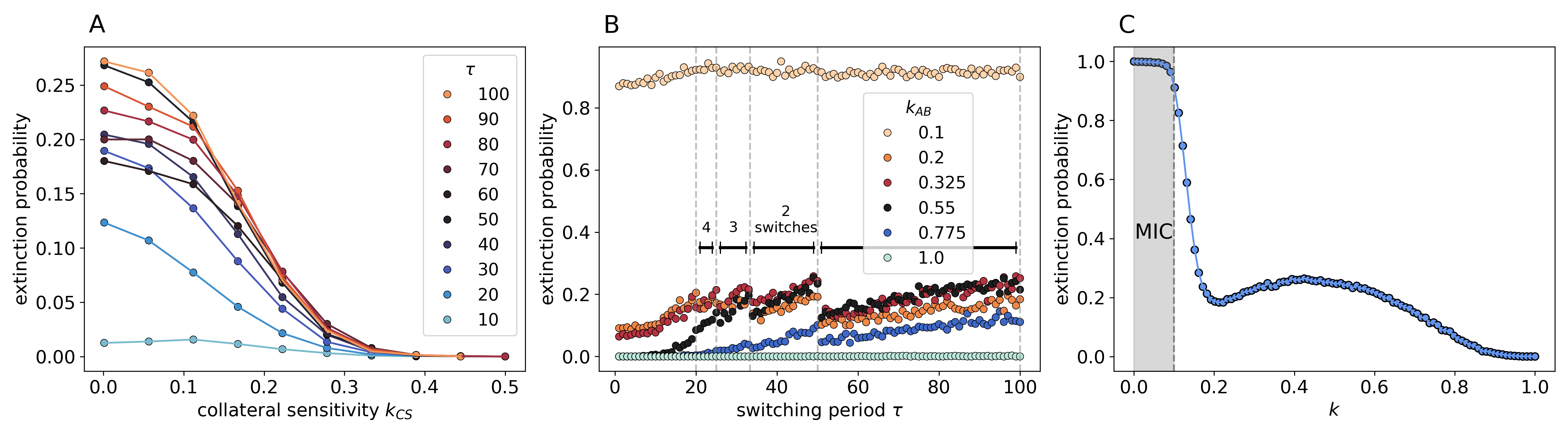}}
\caption[]{Extinction probability sensitivity to model parameters.
(a) Collateral sensitivity (CS) is necessary for extinction. Extinction probabilities decrease as the parameter $k_{CS}$ increases, across a range of switching periods ($\tau$, shown in color).
(b) Increasing antibiotic concentration (lowering $k$) robustly leads to extinction, depending on the number of treatment cycles. We observe a threshold near the MIC, beyond which most populations become extinct regardless of the switching period.
(c) Extinction probability as a function of antibiotic concentration for $\tau = 50$. Close to the MIC, every trajectory becomes extinct. For subinhibitory doses, there is an intermediate dose that maximizes extinction.}
\label{fig:cs}
\end{figure*}

We use the geometric distribution formula to extend these extinction probabilities to various antibiotic changes. However, some care has to be taken with the initial and boundary conditions at the end of the treatment (Appendix~\ref{sec:boundary}). This formula captures our previous intuitions for the two necessary conditions for extinction: the dominance of the single-resistant genotype and the existence of long enough decay times. This simple heuristic matches the shape of extinction probabilities derived from full stochastic simulations (Fig.~\ref{fig:heuristic}(b)) and gives an accurate estimate for cumulative extinction probabilities of populations under one, two, or more antibiotic switches (Fig.~\ref{fig:heuristic}(c)).

\subsection{Additional scenarios: weak collateral sensitivity, increasing antibiotic concentrations, and changing mutation rates}

Our results so far have dealt with strong reciprocal CS, $k_{\rm  CS}=0.05$. If the strength of CS diminishes ($k_{\rm  CS}$ grows), the extinction probabilities at the end of the treatment decrease, as expected given the previous discussion (Fig.~\ref{fig:cs}(a)). Although the dependence of extinction probability on $\tau$ is qualitatively similar, i.e. optimal $\tau$ remain the same throughout, the maximum extinction probability decreases monotonically as $k_{\rm  CS}$ increases, and for $k_{\rm  CS} > 0.3$ it is approximately zero for all $\tau$. That is, for subinhibitory concentrations, CS is a necessary condition for the success of sequential therapies.

We wondered then whether this dependency on CS would be weakened if we increased antibiotic doses. We reasoned that, as $k_A, k_B \to 0$, the total extinction probability should increase. For concentrations close to the minimum inhibitory concentration (MIC), i.e. $k=0.1$ or lower, a threshold behavior emerges where populations go extinct regardless of the switching period (Fig.~\ref{fig:cs}(b,c)), suggesting that extinction is driven primarily by the strength of inhibition rather than the switching dynamics. Indeed, this behavior persists even in the absence of CS (Appendix~\ref{sec:supfigs}, Fig.~\ref{fig:supp-CS-high-dosis}). However, for lower antibiotic concentrations $k>0.1$ we observe a unimodal dependency of the extinction probability on the dose: for a fixed $\tau$, extinction probabilities go up as we decrease $k$ from $1$, and then decrease after a certain dose (Fig.~\ref{fig:cs}(b,c)). We reason that this is due to a change in population dynamics: when antibiotic inhibition increases, the single-resistant mutant quickly dominates the population, which then becomes extinct more easily as we switch the antibiotic. However, if antibiotic inhibition crosses a given threshold, the wild-type population is so low that the influx of resistant mutants actually becomes slower, and therefore the population does not become extinct after switching. We support this qualitative reasoning looking at the mean population before the first switch (Appendix~\ref{sec:supfigs}, Fig.~\ref{fig:supp-unimodalK}).
In the same way, extinction probabilities show a unimodal dependency on mutation: as $\mu_1$ increases, extinction probabilities go up, again due to an increase in the abundance of single-resistant mutants (Appendix~\ref{sec:supfigs}, Fig.~\ref{fig:supp-mutacion}). However, after a critical threshold, the extinction probability starts to decrease, in this case because the double-resistant strain emerges in the population, while the susceptible strain concurrently exhibits a recovery (Appendix~\ref{sec:supfigs}, Fig.~\ref{fig:supp-popratios}). Moreover, as $\mu_2 / \mu_1$ increases, the extinction probability decreases more markedly, which is consistent with the accelerated recovery of the susceptible strain under these conditions, since single-resistant mutants become less frequent when $\mu_2$ is relatively larger.  

\subsection{Optimal switching periods}

After providing an evolutionary population-dynamics argument for the success of sequential therapies based on strong reciprocal CS, and studying the effect of varying several parameters, we return to finding the optimal switching periods.

\begin{figure*}[t] 
  \centerline{\includegraphics[angle=0, width=0.90\textwidth]{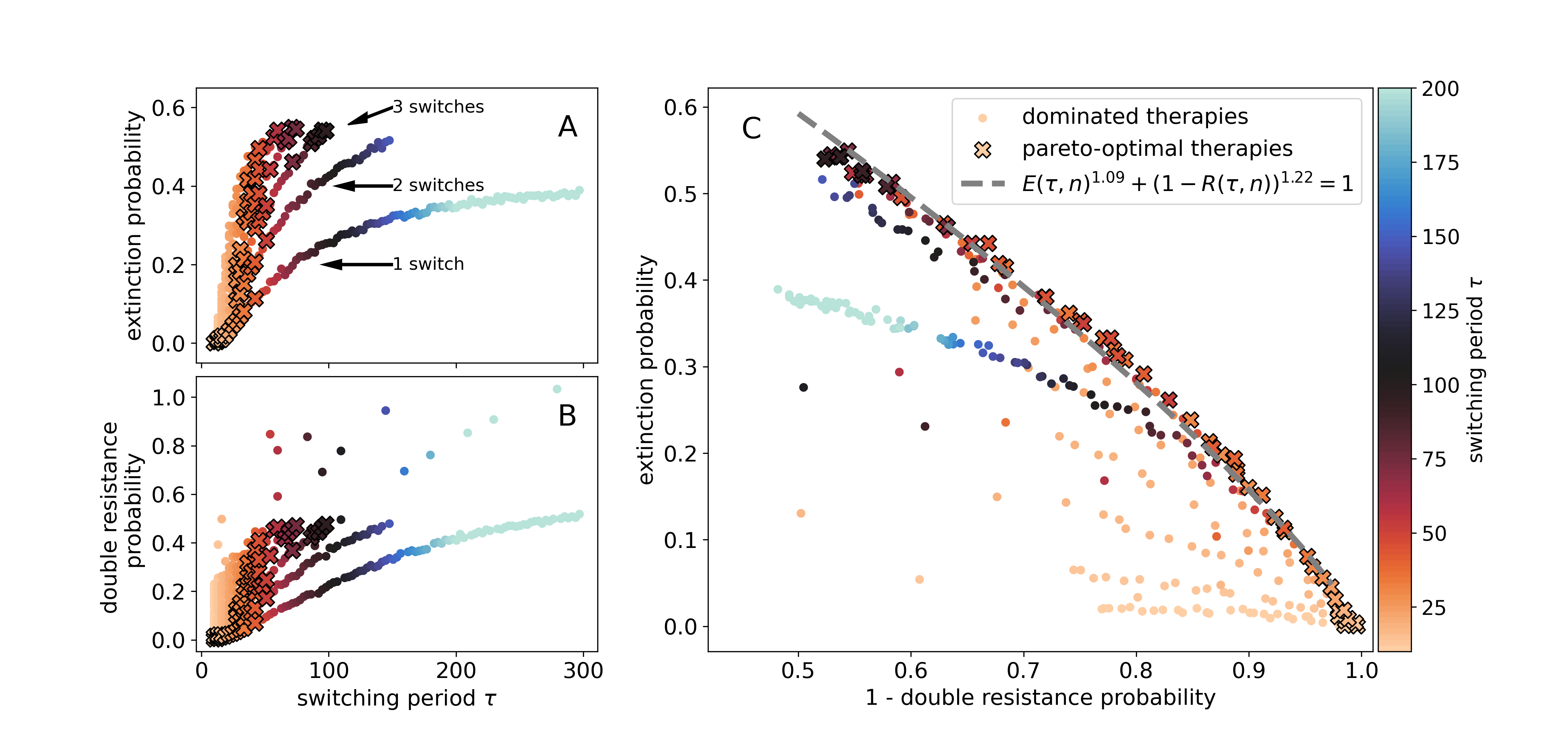}}
\caption[]{Optimal therapy.
(a) Extinction probability as a function of the switching period.
(b) Probability of double resistant mutants at the end of the treatment, conditioned on the population not being extinct.
(c) Pareto front of optimal switching therapies. The crosses indicate therapies that belong to the Pareto front, while the circles show dominated (suboptimal) therapies. The grey dashed line indicates the empirical fit. The probability that a therapy is in the Pareto front is maximized for $\tau \approx 42$. Simulated therapy trajectories were generated with different switching periods ($\tau$) and numbers of treatment rounds ($n$), constrained such that $\tau \cdot n + 50 < 350$, number of trajectories $=10,000$, $k=0.5$, $k_{CS}=0.05$.
}
\label{fig:FigFinalTimes}
\end{figure*}

\begin{figure*}[t] 
  \centerline{\includegraphics[angle=0, width=0.90\textwidth]{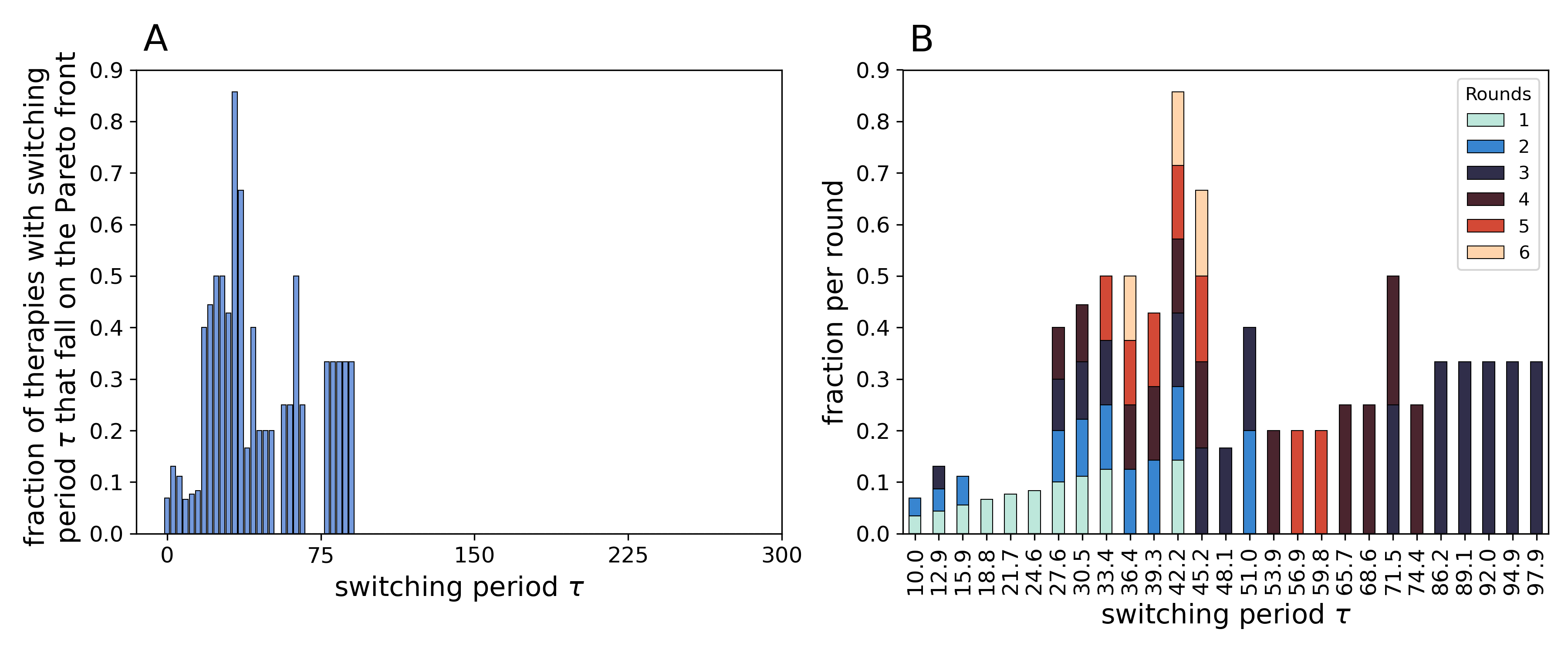}}
\caption[]{Optimal window. Simulated therapy trajectories were generated with different
    switching periods ($\tau$) and numbers of treatment rounds ($n$), constrained such that
    $\tau \cdot n + 50 < 350$.
    (a) Fraction of therapies with a given switching period $\tau$ that lie on the Pareto front.
    (b) Distribution of these fractions across the different numbers of treatment rounds.
    We observe that certain switching periods consistently fall on the Pareto front regardless of the
    number of rounds $\tau \approx 42$, while others are optimal only for one $\tau \approx 20$ or three
    $\tau \approx 90$ switching events. Number of trajectories $=10000$.}
    \label{fig:supp-optimalwindow}
\end{figure*}

So far, we have discussed optimality only in terms of maximizing the cumulative extinction probability at the end of the treatment. In mathematical terms, this function is $E(\tau, n)=1-\big[1-p(\tau)\big]^n$ where $p(\tau)$ is an increasing function of $\tau$, as we have discussed (Fig.~\ref{fig:panel2}(e)). Therefore, $E(\tau,n)$ increases both with $\tau$ and with the number of switches $n$: if the population does not become extinct after the first change, it may do so after subsequent changes. In fact, for a fixed $\tau$, the probability of extinction can increase as much as we want by increasing $n$ (Fig. \ref{fig:FigFinalTimes}(a)). 


However, this comes with a cost: longer therapies also have a higher probability that the double-resistant genotype $x_3$ appears in the population at the end of the treatment (Fig. \ref{fig:FigFinalTimes}(b)), so if the population is not extinct it will become resistant. Hence, there is a tradeoff between maximizing extinction probabilities and minimizing the probability of double-resistance, leading to a Pareto front of feasible therapies (Fig. \ref{fig:FigFinalTimes}(c)) that optimize both objectives. For a fixed extinction probability $E(\tau,n)$, the Pareto front contains those therapies that minimize the probability of double resistance at the end of the treatment, conditioned on the population not being extinct, which we define as $R(\tau,n)=\Pr(N_3(T)\geq 1 | N(T)>0)$. If a therapy is not on the Pareto front, we can always modify it so that either $E(\tau, n)$ or $1-R(\tau, n)$ are improved: they are \emph{dominated} by other, better therapies. Crucially, therapies on the Pareto front cannot improve $E(\tau, n)$ without increasing $R(\tau, n)$ as well. We numerically find that the Pareto front is captured by the curve $E(\tau, n)^{1.09}+(1-R(\tau, n))^{1.22}=1$ (Fig. \ref{fig:FigFinalTimes}(c)).
We can also show that the probability of being in the Pareto front is very high for $\tau \approx 42$ (Fig.~\ref{fig:supp-optimalwindow}(a)), independently of the number of changes (Fig.~\ref{fig:supp-optimalwindow}(b)) and lower everywhere else. Hence, for the parameter set studied in this paper, choosing $\tau$ around that value will lead to optimal therapies: we can then choose whether we want to maximize extinction or minimize resistance by changing the number of switches $n$.


\section{Discussion}

The use of collateral sensitivity (CS) to design sequential therapies has received widespread attention in recent years \cite{imamovic2018drug, sanz2023translating}. The underlying rationale is that antibiotics can be used to steer populations toward genotypic states exhibiting CS, thereby increasing the efficacy of subsequent treatments \cite{nichol2015steering}. In this work, we develop a simple mathematical framework that captures key features of sequential therapies, enabling quantitative exploration of extinction dynamics and providing support for prior evolutionary hypotheses. While experimental studies have demonstrated the potential of sequential therapies to suppress resistance \cite{fuentes2015using, roemhild2018cellular, batra2021high}, theoretical efforts have largely relied on deterministic models where extinctions do not occur \cite{beardmore2010rotating, beardmore2010antibiotic, nyhoegen2023sequential}, although a recent work has used stochastic modeling to explore the effect of antibiotic pulses \cite{morsky2022suppressing}. In contrast to deterministic models, our stochastic modeling framework captures extinction dynamics directly. This enables the identification of optimal treatment strategies based on true eradication events and allows us to explore how extinction probabilities depend on switching timing, antibiotic potency, and mutation rates.


In \cite{beardmore2010rotating}, Beardmore and Peña-Miller state that a successful switching strategy, based on clinical observations \cite{allegranzi2002impact}, is ``\textit{if the observed level of resistance to an antibiotic is too high, exchange it for a different antibiotic}", which is fully consistent with our results. In fact, a key result from our analysis is that fast sequential therapies are suboptimal, and that we need to allow for the evolution of resistance above a threshold in order to eradicate the population after the switch. The tension between waiting long enough for the single-resistant mutant to dominate the population and maximizing the number of switches leads to different optimal switching periods depending on the treatment duration (Fig.~\ref{fig:FigFinalTimes}): longer therapies maximize extinction, but they also cause higher resistance, leading to a Pareto front of optimal switching periods. This gives us some additional insight that will help to choose optimal therapies.

Furthermore, our framework demonstrates that while the requirement for strong CS is a necessary condition for causing extinctions, it allows sequential therapies to outperform traditional monotherapy. In regimes where the antibiotic dose is high, extinction occurs reliably regardless of the switching period (Fig. \ref{fig:cs}B). However, the true strength of this approach lies in its efficacy under subinhibitory concentrations, conditions that are clinically inevitable due to drug diffusion limits within human tissues \cite{muller2004issues}. In these sub-optimal dosage zones, where no-switching treatments fail, our model shows that an optimized sequential protocol can still drive the population to extinction.

In addition, we find a unimodal dose-extinction relationship (Fig.~\ref{fig:cs}(c)), which has been observed before in both experiments \cite{reding2021antibiotic} and models \cite{Czuppon2023} and which can be fully explained using population dynamics arguments: we need single-resistant mutants to dominate the population; these are selected as antibiotic concentration increases, but selection can be hampered if doses are too high. We observe a similar relationship with mutation rates (Appendix~\ref{sec:supfigs}, Fig.~\ref{fig:supp-mutacion}), also supported by population dynamics arguments. This result suggests that increasing mutation rates might be a good complement to this kind of therapies, up to a point where we start facilitating the evolution of double-resistants. This suggestive strategy should be explored with care and checked experimentally before making any therapy recommendations.

This study adopts a deliberately minimal within-host model, four genotypes and two antibiotics with symmetric effects, to isolate the mechanisms of CS-guided switching. Extensions to richer collateral-sensitivity networks \cite{imamovic2018drug, podnecky2018conserved, hernando2020rapid} and to heterogeneous pharmacokinetics/pharmacodynamics \cite{nyhoegen2023sequential} fit naturally within the same framework. CS relationships may vary during treatment \cite{yoshida2017time,barbosa2019evolutionary, sorum2022evolutionary}; such time-dependence can be incorporated by allowing parameters to evolve between switches and calibrating them from data. Empirical evaluation can proceed \textit{in vitro} without altering the theoretical structure. A suitable test system is the ciprofloxacin–tobramycin pair in \textit{P. aeruginosa}, where CS has been observed robustly, and double resistance is infrequent \cite{hernando2020rapid, hernando2023rapid, hernando2025ciprofloxacin}, enabling direct assessment of the model’s predictions.

Our model admits several avenues for generalization beyond the present focus. Beyond bacteria, the mathematical structure of our framework 
maps naturally onto cancer, where CS has been documented across multiple tumor types and treatment classes~\cite{Pluchino2012,
Dhawan2017, Zhao2016, Loria2022}. Sequential exploitation of these
tradeoffs has been
proposed as a strategy to steer tumor clonal evolution toward more vulnerable
states~\cite{Zhao2016}, and experimental evidence using cellular barcoding
has identified CS drugs following resistance to
chemotherapy in colorectal cancer cell lines~\cite{Danisik2023}. More
broadly, the adaptive therapy framework~\cite{Gatenby2009, West2020}
exploits analogous competitive dynamics between sensitive and resistant
subpopulations under modulated treatment schedules, and has been extended to
multidrug sequential protocols~\cite{West2020} that share the same underlying
logic as our model.

Although our focus is on short-term antibiotic treatments, our framework could naturally be applied to chronic infections, where treatment timescales are prolonged and resistance fixation is a central clinical problem. Indeed, sequential therapies have been proposed to treat chronic \textit{Pseudomonas} lung infections where biofilm formation hinders the effectiveness of monotherapies \cite{rojo2016sequential, Musken2018-tz}.

The model could also be lifted to the epidemiological scale, to account transmission dynamics within a population of patients~\cite{bonhoeffer1997evaluating, angst2021comparing, muetter2024impact}. Such extension would introduce additional characteristic timescales, governed by host-to-host interaction frequencies and the mechanisms of transmission, that are qualitatively distinct from the within-host evolutionary dynamics studied here, and might open the possibility of population-level treatment strategies that are not optimal for the individual.

Beyond periodic schedules, our framework can be extended to nonperiodic and adaptive switching policies as considered in \cite{beardmore2010rotating, beardmore2010antibiotic}. Because we treat therapy as a series of switches, each a fresh chance to clear the infection, the same framework lends itself to control-theoretic optimization of switching sequences, including real-time, patient-specific adjustments informed by bacterial-load measurements.

Overall, these results provide quantitative design rules for sequential therapy guided by collateral sensitivity and mathematical modeling, complementing current strategies against antimicrobial resistance.

\begin{figure*}[t] 
  \centerline{\includegraphics[angle=0, width=1.0\textwidth]{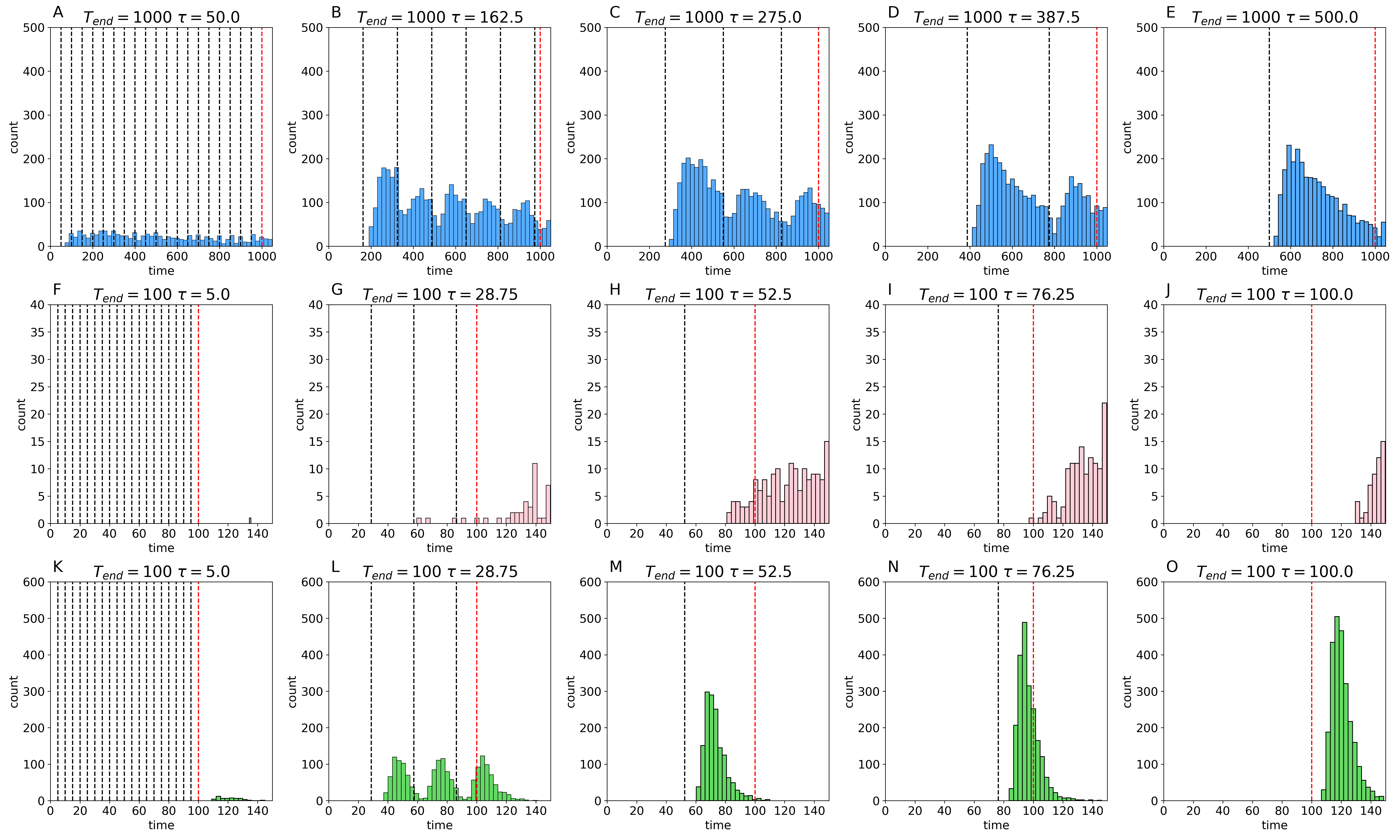}}
\caption[]{Extinction histograms for the hybrid method: Tau-leaping and Gillespie's algorithm.
The top row (blue) shows extinction histograms with a total time of $t_{\text{end treatment}}=1000$ and $t_{\text{end}}=1050$ temporal units. As the antibiotic switching interval increases, extinction events become more clearly clustered after switching events (dashed black vertical lines). Dashed red vertical lines indicate $t_{\text{end treatment}}$. The middle row (pink) presents extinction histograms for a shorter total simulation time of $t_{\text{end treatment}}=100$ and $t_{\text{end}}=150$ temporal units, the same temporal scale used in the main text. At this scale, extinction events appear more diffusely distributed, making it difficult to discern their relationship with antibiotic changes. Both (blue and pink) consider extinction when the number of cells in the population is zero, and switch between tau-leaping and Gillespie's algorithm with a threshold of 10 cells.
The bottom row (green) shows results from Gillespie's algorithm with a threshold of $0.05/\gamma$ for extinction and $15$ cells for switching between tau-leaping and Gillespie, the total simulation time of $t_{\text{end treatment}}=100$ and $t_{\text{end}}=150$ temporal units. The results are consistent with those observed with tau-leaping, supporting the strategy used throughout the main text.}
\label{fig:supp-gillespie}
\end{figure*}


\section*{Acknowledgments}

We thank all members of the Ares-Catalán lab for fruitful discussions, in particular Lucía Yubero for her detailed review of the sigmoid fit. This research was funded by the Spanish Ministerio de Ciencia e Innovaci\'{o}n (MCIN/AEI/10.13039/501100011033) and by ERDF/EU `A way of making Europe' through grant PGE, grant numbers PID2022-142185NB-C21 (S.A.) and PID2022-142185NB-C22 (P.C), grant BASIC, grant number PID2022-141802NB-I00 (J.A.C) and grant GALAR, grant number PID2022-138916NB-I00 (B.P-E.). MICIU/AEI/10.13039/501100011033 has also funded the ‘Severo Ochoa’ Center of Excellence to CNB, CEX2023-001386-S.

\section*{Software availability}
The codes for our simulations, analyses, and production of figures are openly
available~\cite{molina-hernandez2026github}.



\appendix

\section{Justification for the population threshold and tau-leaping approximation}\label{si:gillespie}
In the main text, we considered a stochastic model of bacterial population dynamics under antibiotic treatment, where extinction was defined as the population dropping below a threshold of $0.05\gamma^{-1}=5$ individuals. Simulations were performed using the tau-leaping method to efficiently approximate the stochastic process. Here, we justify this approach by comparing it with a hybrid method that uses tau-leaping for populations with many individuals and Gillespie's algorithm for populations below that limit.

\begin{figure}[t]
  \centering
    \includegraphics[width=0.45\textwidth]{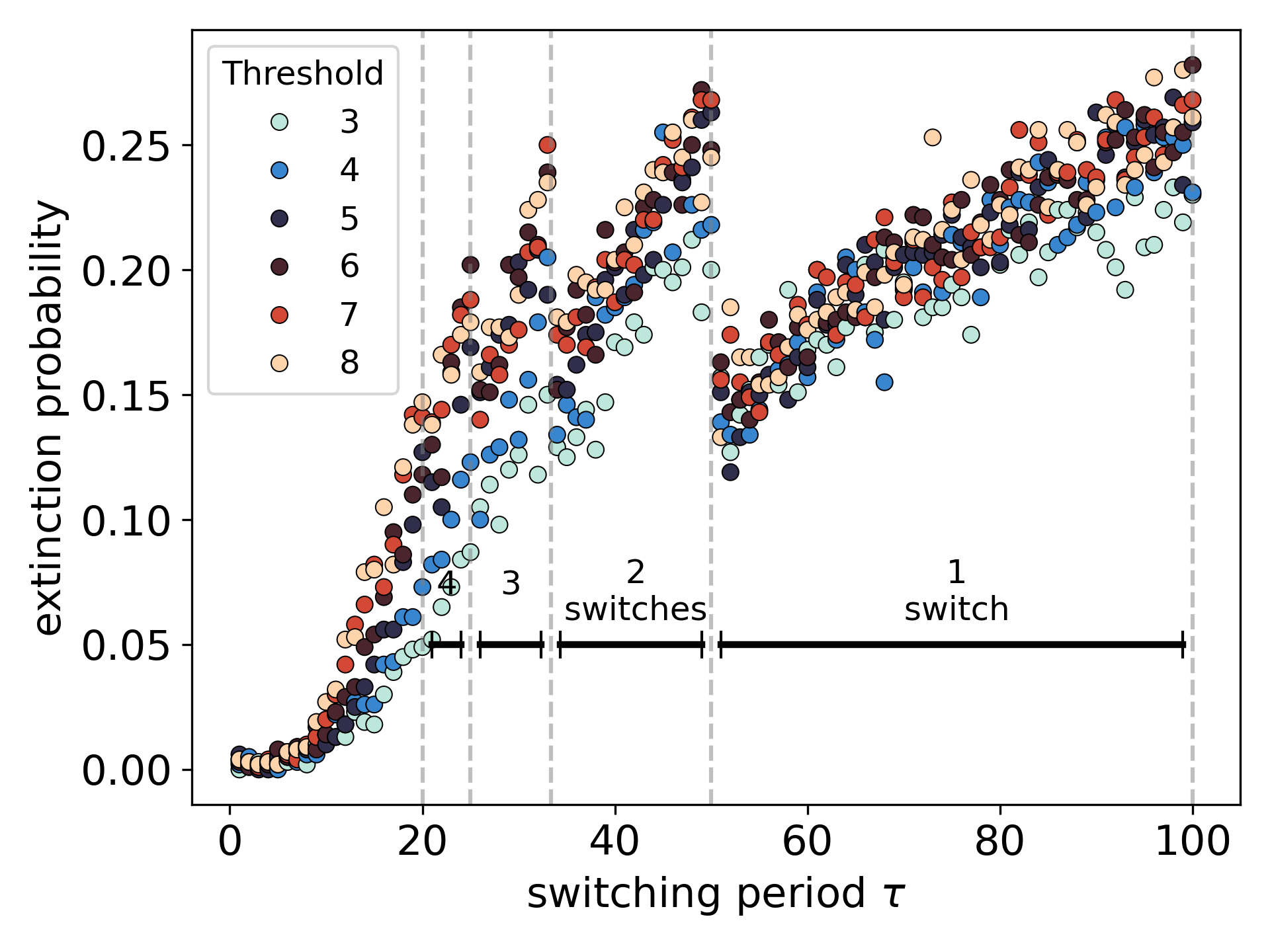}
    \caption{Changing the extinction threshold does not change the qualitative behavior of the extinction curve.
Different thresholds for determining when a population is considered to be extinct, from 3\% to 8\% of theoretical carrying capacity. In the main text 5\% has been used as the threshold.}
\label{fig:supp-thresholds}
\end{figure}

\begin{figure*}[t] 
  \centerline{\includegraphics[angle=0, width=0.6\textwidth]{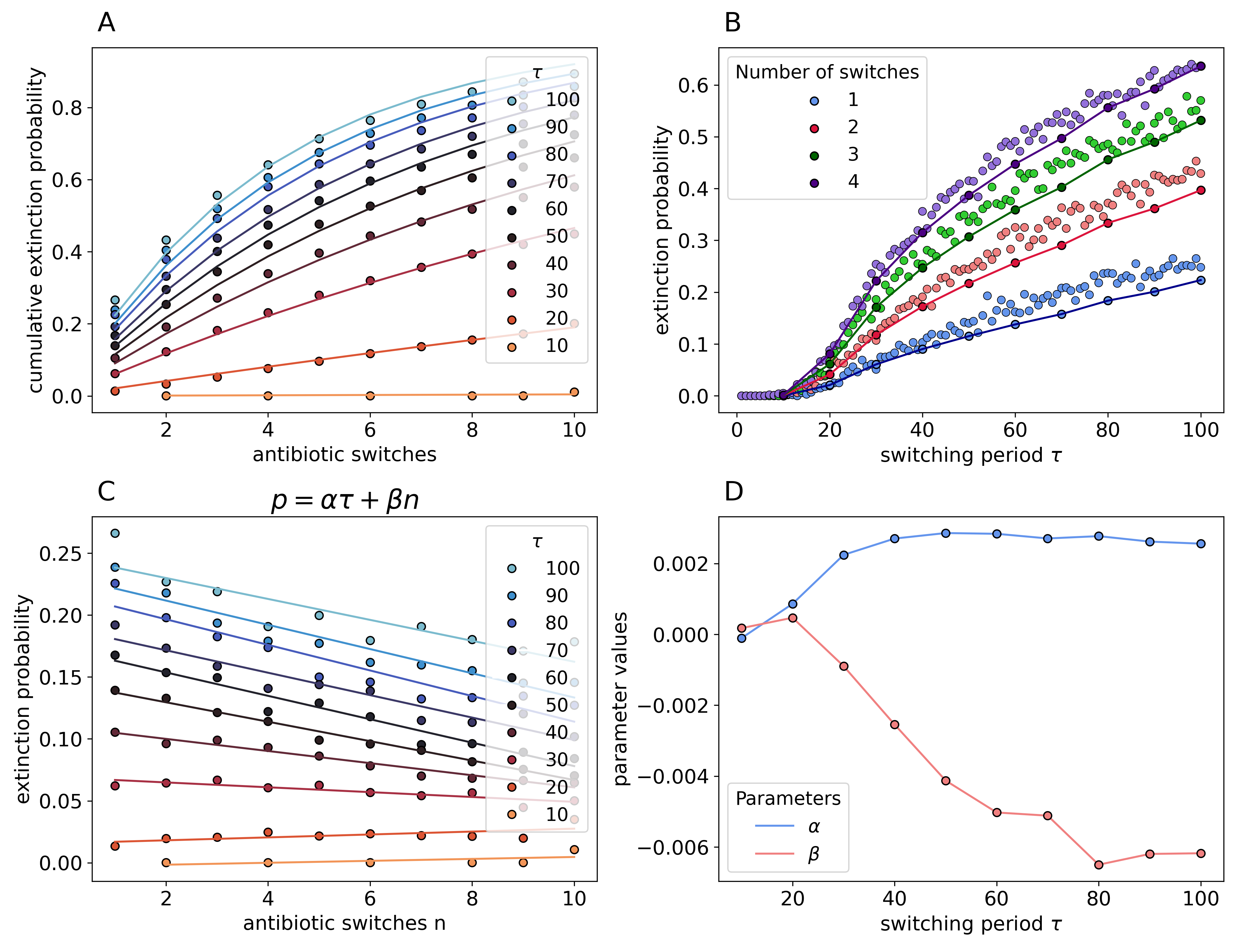}}
\caption[]{ 
(a) Cumulative extinction probability over 10 treatment switches with different switching periods $\tau$. Points represent extinction probabilities estimated through simulation, while the solid line reflects a fit to a geometric distribution $1-(1-p)^n$.
(b) Estimation of the extinction probability for different number of antibiotic switches (represented with colors as indicated in the legend) using the fit of the geometric model obtained in (a).
(c) Linear fit to the extinction probability after each switch for the hierarchical geometric model.
(d) Estimated parameters for each tau $\tau$ in (c).
}
\label{fig:supp-noneq}
\end{figure*}

Gillespie's algorithm provides an exact simulation of the underlying stochastic process but becomes computationally prohibitive when population size is large. Specifically, as the number of bacteria increases, the waiting times between reaction events shorten, leading to a significant increase in computational cost. This has a direct impact on the feasibility of simulating extinction events under different antibiotic switching strategies.

To assess whether the chosen threshold and tau-leaping approximation capture the relevant extinction dynamics, we performed simulations using the hybrid method and compared the results. Our key observation is that extinction events occur primarily after an antibiotic switch, with a characteristic delay. However, when antibiotic switching times are of the same order of magnitude as reaction times at low population sizes, this pattern is no longer evident. Instead, we observe a diffuse cloud of extinction events, making it difficult to distinguish the effect of antibiotic changes. To recover the expected correlation between switching events and extinctions, it is necessary to increase the switching times.

To illustrate these findings, we present extinction histograms from three sets of simulations. The first set (Fig.~\ref{fig:supp-gillespie}, top row) corresponds to a longer total simulation time of 1000 temporal units, allowing for a clearer observation of extinction clustering after antibiotic switches. The second set (Fig.~\ref{fig:supp-gillespie}, middle row) corresponds to a shorter total simulation time of 100 temporal units, matching the temporal scale used in the main text. In this shorter time frame, extinction events appear as a diffuse pattern, making it harder to discern correlations with antibiotic changes. In contrast, with a sufficiently long observation window, the extinction events align with antibiotic switching events, confirming that the primary driver of extinction is the switching strategy itself. In these two sets, we have considered extinctions when we have a population of zero bacteria. We add a third set using the hybrid method but with the same threshold for the extinctions used in the main text (Fig.~\ref{fig:supp-gillespie}, bottom row), confirming that the results are consistent and that the tau-leaping algorithm does not qualitatively alter the observed patterns.

We also compared the distribution of extinction times for fast-extinction events (such as those involving a high antibiotic dose, $\kappa=0.1$) and observed an overlap between the two histograms. This makes the tau-leaping approximation also a good approximation of the solution to the master equation in this regime.

Finally, we tested the robustness of the extinction threshold. In the main text, we consider a population extinct if it goes below 5\% carrying capacity. We varied this threshold between 3\% and 8\% and obtained quantitatively very similar results (Fig.~\ref{fig:supp-thresholds}).

\section{Parameters used in the main text}
\label{app:table1}

Unless otherwise stated, all graphs in the main text were made with the parameter values in Table~\ref{tab:parameters}.

\begin{table*}
\centering
\caption{Model parameters and their meanings}
\begin{tabular}{|c|l|c|}
\hline
\textbf{Parameter} & \textbf{Description} & \textbf{Value} \\
\hline
$\tau$        & Duration of treatment with antibiotic $A$ or $B$ & Variable \\
$k_A$, $k_B$  & Effect of bacteriostatic antibiotics (higher $\Rightarrow$ weaker inhibition) & 0.5 \\
$\gamma$      & Inverse of the theoretical carrying capacity & 0.01 \\
$k_{\rm  CS}$ & Collateral sensitivity effect (higher $\Rightarrow$ weaker CS) & 0.05 \\
$\mu_1$       & Mutation rate to a resistant genotype & $10^{-3}$ \\
$\mu_2$       & Mutation rate back to a sensitive genotype & $10^{-2}$ \\
$t_{\text{end}}$     & Total time of the simulation & 150 \\
$t_{\text{end treatment}}$ & Time in which we allow antibiotics to be switched & 100 \\
\hline
\end{tabular}%
\label{tab:parameters}
\end{table*}

\section{Non-equilibrium effects and a hierarchical geometric model}\label{sec:hierarchical}

\begin{figure}[t] 
  \centerline{\includegraphics[angle=0, width=0.43\textwidth]{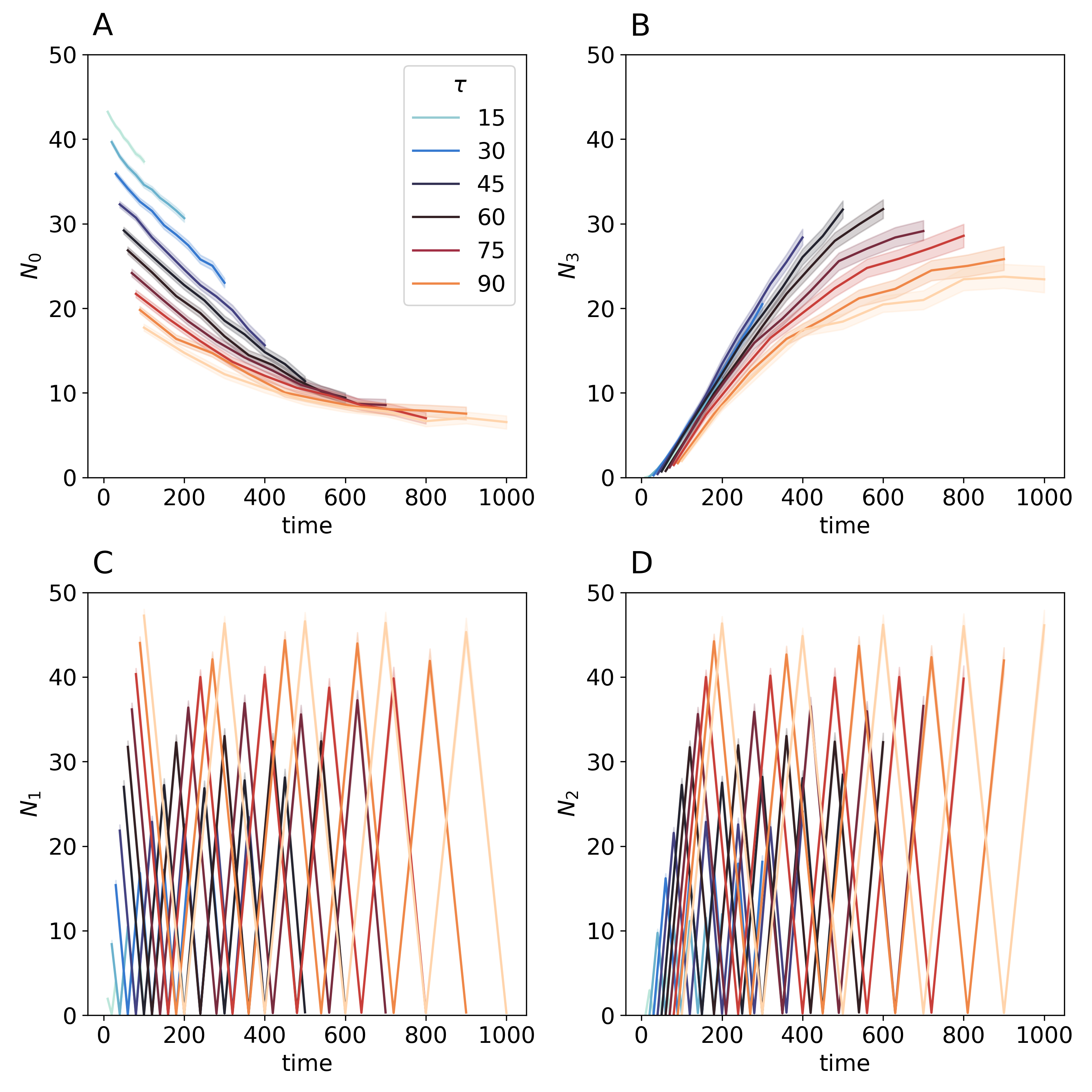}}
\caption{Mean population trajectories. Mean values of the trajectories before the switch simulated for the geometric fit. Shades indicate 95\% confidence interval. We observe that surviving populations achieve double-resistant mutants. 
(a) $N_0$. (b) $N_3$. (c) $N_1$ oscillatory behavior. (d) $N_2$ oscillatory behavior.}
\label{fig:supp-population}
\end{figure}

The discrepancies observed in the geometric model fit from Fig.~\ref{fig:supp-noneq}(a,b) can be explained by the fact that the system is not in equilibrium. To better understand this behavior, we analyzed the extinction probability per round, $p$, as a function of time and the number of rounds. We performed a linear fit of the form $p = \alpha \tau + \beta n$, where $p$ is the extinction probability, $\tau$ is the switching period, and $n$ is the number of completed treatment rounds (Fig.~\ref{fig:supp-noneq}(c)). This model captures the systematic dependence of extinction dynamics on both time and the structure of the treatment.

We then used this time-dependent extinction probability to construct a hierarchical geometric model for the cumulative extinction probability:
\[
P_{\text{ext}} = 1 - (1 - \alpha \tau - \beta n)^n,
\]
which fits the data well over a wide range of conditions (Fig.~\ref{fig:panel2}(d)).

To explore how the fitting parameters vary with the switching period, we examined the dependence of $\alpha$ and $\beta$ on $\tau$ (Fig.~\ref{fig:supp-noneq}(d)). For small values of $\tau$, we find $\beta > 0$, indicating that the probability of extinction increases with the number of rounds. This could be due to the oscillatory regime driving the system into extinction-prone states. In contrast, for large $\tau$, we observe $\beta < 0$, suggesting that prolonged exposure in each cycle may favor the emergence of double-resistant mutants, reducing extinction probability over time (Fig.~\ref{fig:supp-population}). Together, these results highlight the importance of non-equilibrium effects in shaping extinction outcomes during sequential therapy.


\section{Parameters of the sigmoid fit}\label{sec:sigmoid}

We introduce as parameters to be adjusted in a logistic model the population before the change of antibiotic. Depending on the antibiotic we are using, the populations of $x_1$ and $x_2$ have different growth rates, and therefore we consider the variables
$$\begin{cases}
    x_{iA} = 0 \quad \text{if antibiotic is $B$,} \quad x_i \quad \text{otherwise} \\
    x_{iB} = x_i \quad \text{if antibiotic is $B$,} \quad 0 \quad \text{otherwise}
\end{cases}$$

for $i=1,2$. We therefore end up with the following inputs: $x_0$, $x_3$, $x_{1A}$, $x_{1B}$, $x_{2A}$, $x_{2B}$ (Table~\ref{tab:parameterssigmoid1}). 

We observed that having $x_0$ or $x_3$ in the population before the switch reduces the probability of extinction (the parameters are negative). The same happens when we have bacteria that will be resistant after switching, $x_{1B}$ and $x_{2A}$. The best scenario for extinction is when all the population is dominated by either $x_{1A}$ or $x_{2B}$. 
The dataset used for these fits is the population composition before the switch of $10,000$ trajectories per $\tau$ simulated to estimate the extinction probability of Fig.~\ref{fig:panel2}(a).


\begin{table}[h!]
\centering
\caption{Parameters in sigmoid fit of Fig. \ref{fig:heuristic}}
\small 
{
\begin{tabular}{|c|r|}
\hline
\textbf{Parameter} &  \textbf{Sigmoid fit} \\
\hline
$x_0$    & $-0.820103$\phantom{x} \\
$x_3$    & $-1.243756$\phantom{x} \\
$x_{1A}$ & $0.019319$\phantom{x} \\
$x_{1B}$ & $-0.572685$\phantom{x} \\
$x_{2A}$ & $-0.545886$\phantom{x} \\
$x_{2B}$ & $0.013334$\phantom{x} \\
\hline
\end{tabular}
}
\label{tab:parameterssigmoid1}
\end{table}


\begin{figure}[t]
  \centering
    \includegraphics[width=0.45\textwidth]{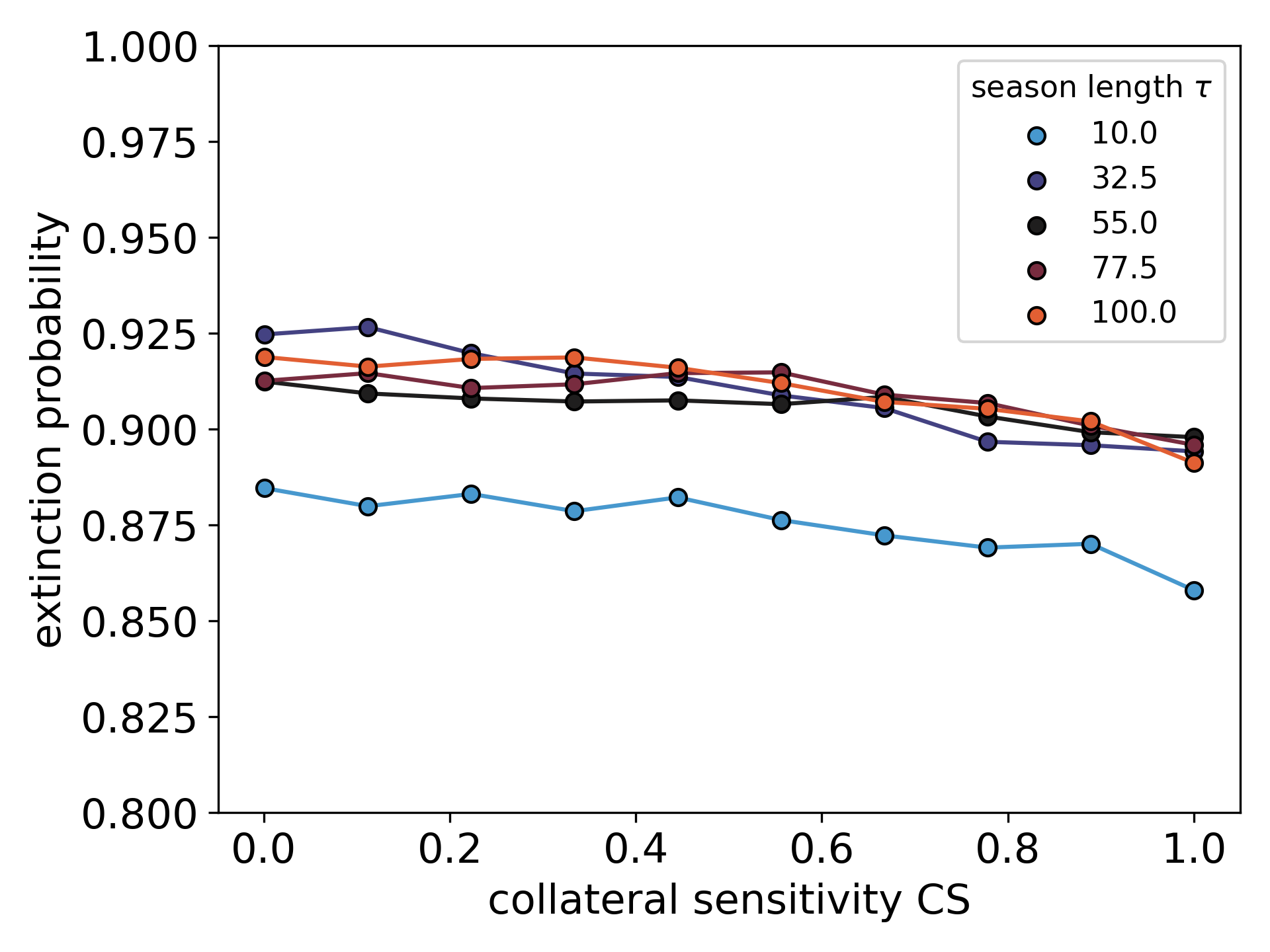}
    \caption{Collateral sensitivity is irrelevant when high doses of antibiotic are applied. Simulations with $k=0.1$.
}
\label{fig:supp-CS-high-dosis}
\end{figure}

\begin{figure}[t]
  \centering
    \includegraphics[width=0.30\textwidth]{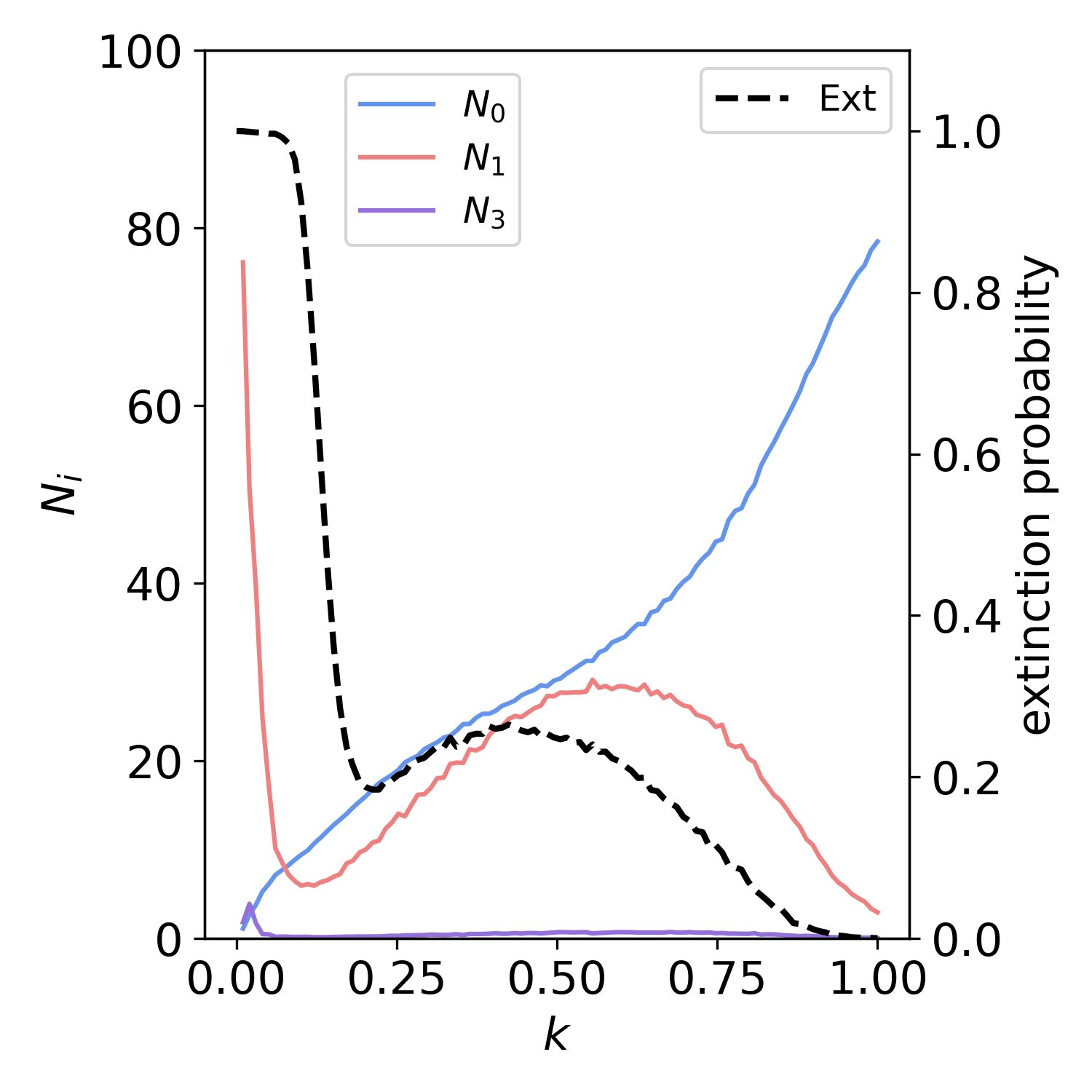}
    \caption{Effect of changes in antibiotic concentration. Mean population composition prior to the first switch (colored lines), superimposed with the extinction probability (black dashed line). As antibiotic inhibition increases ($k$ decreases), the population of $x_1$ goes up, which leads to an increase in extinction probability. After a certain antibiotic concentration (below $k\approx 0.7$), $x_0$ does not reach high values, and evolution to $x_1$ slows down. This leads to a decrease in extinction at intermediate to high antibiotic concentrations ($k < 0.4$ approximately). When antibiotic inhibition becomes very strong (i.e. $k<0.1$), the decrease in $x_0$ is so strong that populations become extinct not due to collateral sensitivity, but because the overall carrying capacity is very low and extinctions occur from fluctuations. Simulations with $\tau=50$, $t_{\text{end}}=150$, $t_{\text{end treatment}}=100$, $\text{number of trajectories}=10000$.
}
\label{fig:supp-unimodalK}
\end{figure}

\begin{figure}[t]
  \centering
    \includegraphics[width=0.48\textwidth]{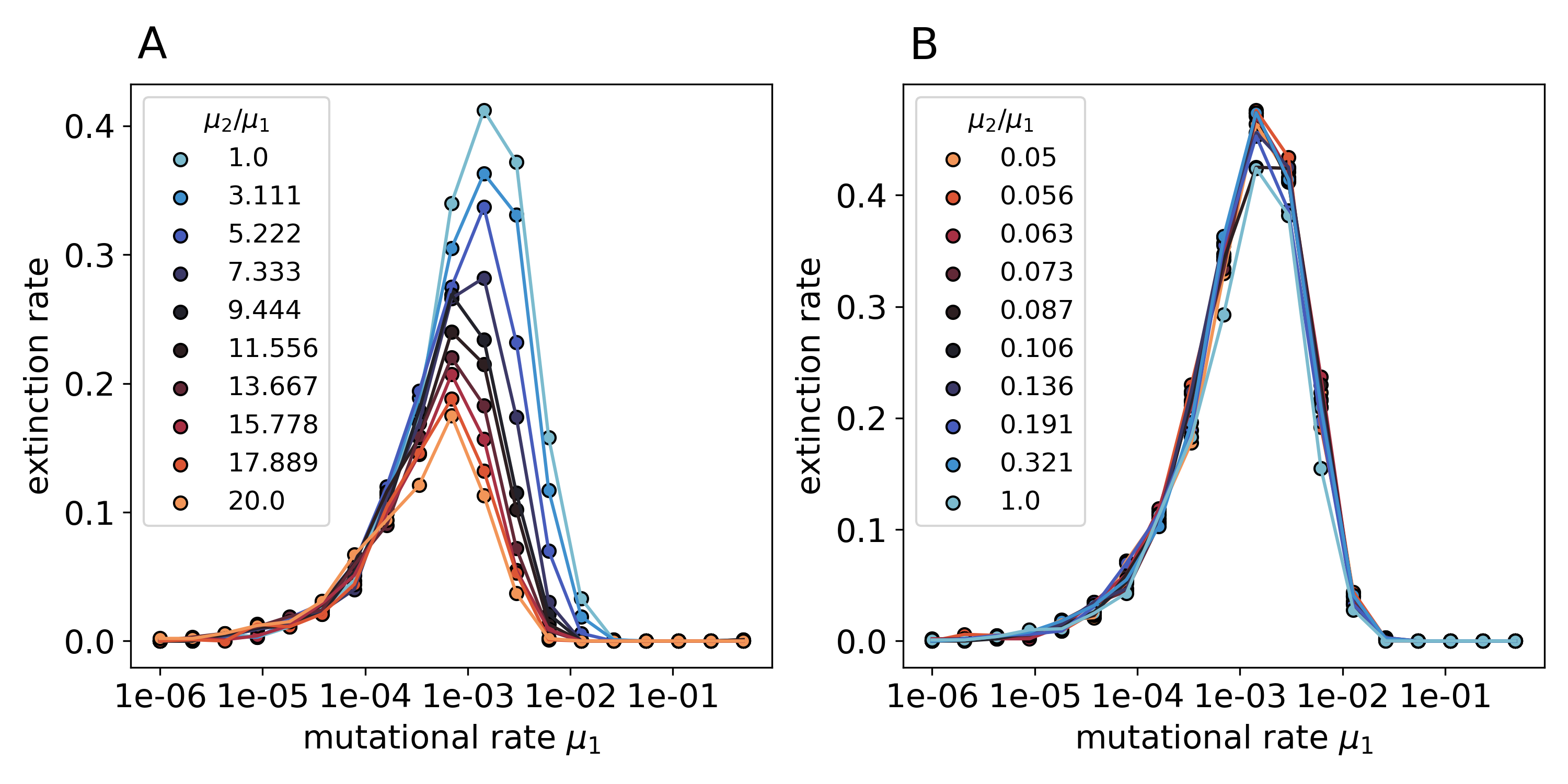}
\caption{Effect of changes in mutation rates.
(a) Extinction rate as a function of mutation rate $\mu_1$, for a range
of ratios $\mu_2/\mu_1$ ranging from 1 to 20 (see legend), i.e.\
$\mu_2 \geq \mu_1$. (b) Same as (a) but the ratio $\mu_2/\mu_1$ now
goes from 0.05 to 1 (see legend), i.e.\ $\mu_1 > \mu_2$. The
extinction probability exhibits a nonmonotonic dependence on the
baseline mutation rate ($\mu_1$), peaking at intermediate values. The
ratio between mutation rates $\mu_2/\mu_1$ modulates the peak height
and position, but does not otherwise alter this behavior. Simulations
with $\tau=50$, $t_{\rm end}=150$, $t_{\rm end\ treatment}=100$,
number of trajectories $=1000$.
}
\label{fig:supp-mutacion}
\end{figure}

\begin{figure*}[t] 
  \centerline{\includegraphics[angle=0, width=0.55\textwidth]{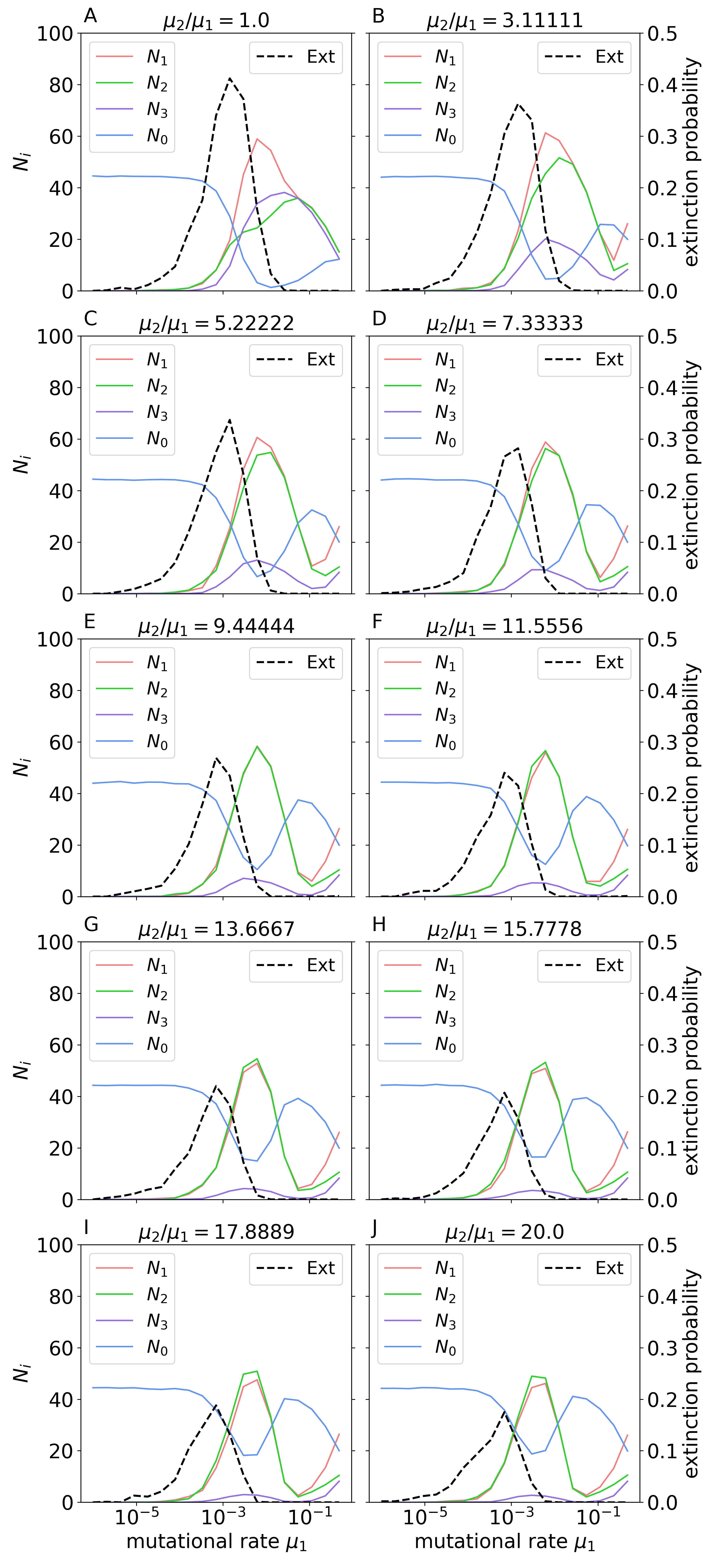}}
\caption[]{ Effect of changes in mutation rates. Mean population composition prior to the switch for trajectories that do not become extinct (colored lines), and extinction probability (black dashed line). The decline of the extinction peak coincides with the emergence of the double-resistant strain and with the recovery of the susceptible strain, the latter becoming more pronounced as $\mu_2$ increases relative to $\mu_1$. $\tau=50$, $t_{\text{end}}=150$, $t_{\text{end treatment}}=100$, $\text{number of trajectories}=1000$.
}
\label{fig:supp-popratios}
\end{figure*}

\section{Formula for the extension of the heuristic}\label{sec:boundary}

Due to the boundary and initial conditions, Eq.~\eqref{eq:heuristica} in the main text cannot be introduced directly into the geometric distribution for extending the extinction probability to multiple switches. Remember that $p(\tau)$ has two contributions, $p_r(\tau)$, the probability that the single-resistant mutant is not completely dominant in the population, and $p_d(\tau)$, the probability that a population gets extinct in a time smaller than $\tau$.


For the first antibiotic switch $p_{\mathrm{init}}(\tau)$, we calculate $p_{r, \mathrm{init}}(\tau)$ as
\[
    p_{r, \mathrm{init}}(\tau) = \text{Pr}[x_0+x_2+x_3 \geq 2 |\tau, x_0(0)=50]
\]
Similarly, after the last antibiotic switch, $p_{\mathrm{bound}}(\tau)$,  we modify the decay probability to $p_{d, \mathrm{bound}}=p_d(t_{\rm end}  \text{ (mod } \tau + 50) )$, because the population has more time to decay, as a consequence of the chosen boundary conditions.

Putting everything together, we have the formula
\[
p_{\mathrm{ext}}(\tau) = 1 - [1-p(\tau)]^{n-2}[1-p_{\mathrm{init}}(\tau)][1-p_{\mathrm{bound}}(\tau)].
\]

\section{Supplementary figures}\label{sec:supfigs}

Figs.~\ref{fig:supp-CS-high-dosis}--\ref{fig:supp-popratios} contain supplementary information referred from the main text.

\clearpage

\bibstyle{unsrt}
\bibliography{bibliography}

@article{naghavi2024global,
  title={Global burden of bacterial antimicrobial resistance 1990--2021: a systematic analysis with forecasts to 2050},
  author={Naghavi, Mohsen and Vollset, Stein Emil and Ikuta, Kevin S and Swetschinski, Lucien R and Gray, Authia P and Wool, Eve E and Aguilar, Gisela Robles and Mestrovic, Tomislav and Smith, Georgia and Han, Chieh and others},
  journal={The Lancet},
  volume={404},
  number={10459},
  pages={1199--1226},
  year={2024},
  publisher={Elsevier}
}

@article{catalan2022seeking,
  title={Seeking patterns of antibiotic resistance in ATLAS, an open, raw MIC database with patient metadata},
  author={Catal{\'a}n, Pablo and Wood, Emily and Blair, Jessica MA and Gudelj, Ivana and Iredell, Jonathan R and Beardmore, Robert E},
  journal={Nature Communications},
  volume={13},
  number={1},
  pages={2917},
  year={2022},
  publisher={Nature Publishing Group UK London}
}

@article{roemhild2022physiology,
  title={The physiology and genetics of bacterial responses to antibiotic combinations},
  author={Roemhild, Roderich and Bollenbach, Tobias and Andersson, Dan I},
  journal={Nature Reviews Microbiology},
  volume={20},
  number={8},
  pages={478--490},
  year={2022},
  publisher={Nature Publishing Group UK London}
}

@article{tyers2019drug,
  title={Drug combinations: a strategy to extend the life of antibiotics in the 21st century},
  author={Tyers, Mike and Wright, Gerard D},
  journal={Nature Reviews Microbiology},
  volume={17},
  number={3},
  pages={141--155},
  year={2019},
  publisher={Nature Publishing Group UK London}
}

@article{kavvcivc2020mechanisms,
  title={Mechanisms of drug interactions between translation-inhibiting antibiotics},
  author={Kav{\v{c}}i{\v{c}}, Bor and Tka{\v{c}}ik, Ga{\v{s}}per and Bollenbach, Tobias},
  journal={Nature Communications},
  volume={11},
  number={1},
  pages={4013},
  year={2020},
  publisher={Nature Publishing Group UK London}
}

@article{rosenkilde2019collateral,
  title={Collateral sensitivity constrains resistance evolution of the CTX-M-15 $\beta$-lactamase},
  author={Rosenkilde, Carola EH and Munck, Christian and Porse, Andreas and Linkevicius, Marius and Andersson, Dan I and Sommer, Morten OA},
  journal={Nature Communications},
  volume={10},
  number={1},
  pages={618},
  year={2019},
  publisher={Nature Publishing Group UK London}
}

@article{baym2016multidrug,
  title={Multidrug evolutionary strategies to reverse antibiotic resistance},
  author={Baym, Michael and Stone, Laura K and Kishony, Roy},
  journal={Science},
  volume={351},
  number={6268},
  pages={aad3292},
  year={2016},
  publisher={American Association for the Advancement of Science}
}

@article{tamma2012combination,
  title={Combination therapy for treatment of infections with gram-negative bacteria},
  author={Tamma, Pranita D and Cosgrove, Sara E and Maragakis, Lisa L},
  journal={Clinical Microbiology Reviews},
  volume={25},
  number={3},
  pages={450--470},
  year={2012},
  publisher={Am Soc Microbiol}
}

@article{pena2013most,
  title={When the most potent combination of antibiotics selects for the greatest bacterial load: the smile-frown transition},
  author={Pena-Miller, Rafael and Laehnemann, David and Jansen, Gunther and Fuentes-Hernandez, Ayari and Rosenstiel, Philip and Schulenburg, Hinrich and Beardmore, Robert},
  journal={PLoS Biology},
  volume={11},
  number={4},
  pages={e1001540},
  year={2013},
  publisher={Public Library of Science San Francisco, USA}
}

@article{vestergaard2016antibiotic,
  title={Antibiotic combination therapy can select for broad-spectrum multidrug resistance in Pseudomonas aeruginosa},
  author={Vestergaard, Martin and Paulander, Wilhelm and Marvig, Rasmus L and Clasen, Julie and Jochumsen, Nicholas and Molin, S{\o}ren and Jelsbak, Lars and Ingmer, Hanne and Folkesson, Anders},
  journal={International Journal of Antimicrobial Agents},
  volume={47},
  number={1},
  pages={48--55},
  year={2016},
  publisher={Elsevier}
}

@article{roemhild2019evolutionary,
  title={Evolutionary ecology meets the antibiotic crisis: Can we control pathogen adaptation through sequential therapy?},
  author={Roemhild, Roderich and Schulenburg, Hinrich},
  journal={Evolution, Medicine, and Public Health},
  volume={2019},
  number={1},
  pages={37--45},
  year={2019},
  publisher={Oxford University Press}
}

@article{beardmore2017antibiotic,
  title={Antibiotic cycling and antibiotic mixing: which one best mitigates antibiotic resistance?},
  author={Beardmore, Robert Eric and Pe{\~n}a-Miller, Rafael and Gori, Fabio and Iredell, Jonathan},
  journal={Molecular Biology and Evolution},
  volume={34},
  number={4},
  pages={802--817},
  year={2017},
  publisher={Oxford University Press}
}

@article{barbosa2017alternative,
  title={Alternative evolutionary paths to bacterial antibiotic resistance cause distinct collateral effects},
  author={Barbosa, Camilo and Trebosc, Vincent and Kemmer, Christian and Rosenstiel, Philip and Beardmore, Robert and Schulenburg, Hinrich and Jansen, Gunther},
  journal={Molecular Biology and Evolution},
  volume={34},
  number={9},
  pages={2229--2244},
  year={2017},
  publisher={Oxford University Press}
}

@article{imamovic2018drug,
  title={Drug-driven phenotypic convergence supports rational treatment strategies of chronic infections},
  author={Imamovic, Lejla and Ellabaan, Mostafa Mostafa Hashim and Machado, Ana Manuel Dantas and Citterio, Linda and Wulff, Tune and Molin, Soren and Johansen, Helle Krogh and Sommer, Morten Otto Alexander},
  journal={Cell},
  volume={172},
  number={1},
  pages={121--134},
  year={2018},
  publisher={Elsevier}
}

@article{hernando2020rapid,
  title={Rapid and robust evolution of collateral sensitivity in Pseudomonas aeruginosa antibiotic-resistant mutants},
  author={Hernando-Amado, Sara and Sanz-Garc{\'\i}a, Fernando and Mart{\'\i}nez, Jos{\'e} Luis},
  journal={Science Advances},
  volume={6},
  number={32},
  pages={eaba5493},
  year={2020},
  publisher={American Association for the Advancement of Science}
}

@article{podnecky2018conserved,
  title={Conserved collateral antibiotic susceptibility networks in diverse clinical strains of Escherichia coli},
  author={Podnecky, Nicole L and Fredheim, Elizabeth GA and Kloos, Julia and S{\o}rum, Vidar and Primicerio, Raul and Roberts, Adam P and Rozen, Daniel E and Samuelsen, {\O}rjan and Johnsen, P{\aa}l J},
  journal={Nature Communications},
  volume={9},
  number={1},
  pages={3673},
  year={2018},
  publisher={Nature Publishing Group UK London}
}

@article{fuentes2015using,
  title={Using a sequential regimen to eliminate bacteria at sublethal antibiotic dosages},
  author={Fuentes-Hernandez, Ayari and Plucain, Jessica and Gori, Fabio and Pena-Miller, Rafael and Reding, Carlos and Jansen, Gunther and Schulenburg, Hinrich and Gudelj, Ivana and Beardmore, Robert},
  journal={PLoS Biology},
  volume={13},
  number={4},
  pages={e1002104},
  year={2015},
  publisher={Public Library of Science San Francisco, CA USA}
}

@article{batra2021high,
  title={High potency of sequential therapy with only $\beta$-lactam antibiotics},
  author={Batra, Aditi and Roemhild, Roderich and Rousseau, Emilie and Franzenburg, S{\"o}ren and Niemann, Stefan and Schulenburg, Hinrich},
  journal={eLife},
  volume={10},
  pages={e68876},
  year={2021},
  publisher={eLife Sciences Publications, Ltd}
}

@article{maltas2025dynamic,
  title={Dynamic collateral sensitivity profiles highlight opportunities and challenges for optimizing antibiotic treatments},
  author={Maltas, Jeff and Huynh, Anh and Wood, Kevin B},
  journal={PLoS Biology},
  volume={23},
  number={1},
  pages={e3002970},
  year={2025},
  publisher={Public Library of Science San Francisco, CA USA}
}

@article{maltas2019pervasive,
  title={Pervasive and diverse collateral sensitivity profiles inform optimal strategies to limit antibiotic resistance},
  author={Maltas, Jeff and Wood, Kevin B},
  journal={PLoS Biology},
  volume={17},
  number={10},
  pages={e3000515},
  year={2019},
  publisher={Public Library of Science San Francisco, CA USA}
}

@article{weaver2024reinforcement,
  title={Reinforcement learning informs optimal treatment strategies to limit antibiotic resistance},
  author={Weaver, Davis T and King, Eshan S and Maltas, Jeff and Scott, Jacob G},
  journal={Proceedings of the National Academy of Sciences of the U.S.A.},
  volume={121},
  number={16},
  pages={e2303165121},
  year={2024},
  publisher={National Acad Sciences}
}

@article{aulin2021design,
  title={Design principles of collateral sensitivity-based dosing strategies},
  author={Aulin, Linda BS and Liakopoulos, Apostolos and van der Graaf, Piet H and Rozen, Daniel E and van Hasselt, JG Coen},
  journal={Nature Communications},
  volume={12},
  number={1},
  pages={5691},
  year={2021},
  publisher={Nature Publishing Group UK London}
}

@article{gillespie1976general,
  title={A general method for numerically simulating the stochastic time evolution of coupled chemical reactions},
  author={Gillespie, Daniel T},
  journal={Journal of Computational Physics},
  volume={22},
  number={4},
  pages={403--434},
  year={1976},
  publisher={Elsevier}
}

@article{gillespie2007stochastic,
  title={Stochastic simulation of chemical kinetics},
  author={Gillespie, Daniel T},
  journal={Annual Review of Physical Chemistry},
  volume={58},
  number={1},
  pages={35--55},
  year={2007},
  publisher={Annual Reviews}
}

@article{reding2021antibiotic,
  title={The Antibiotic Dosage of Fastest Resistance Evolution: gene amplifications underpinning the inverted-U},
  author={Reding, Carlos and Catal{\'a}n, Pablo and Jansen, Gunther and Bergmiller, Tobias and Wood, Emily and Rosenstiel, Phillip and Schulenburg, Hinrich and Gudelj, Ivana and Beardmore, Robert},
  journal={Molecular Biology and Evolution},
  volume={38},
  number={9},
  pages={3847--3863},
  year={2021},
  publisher={Oxford University Press}
}

@article{andersson2012evolution,
  title={Evolution of antibiotic resistance at non-lethal drug concentrations},
  author={Andersson, Dan I and Hughes, Diarmaid},
  journal={Drug Resistance Updates},
  volume={15},
  number={3},
  pages={162--172},
  year={2012},
  publisher={Elsevier}
}

@article{angst2021comparing,
  title={Comparing treatment strategies to reduce antibiotic resistance in an in vitro epidemiological setting},
  author={Angst, Daniel C and Tepekule, Burcu and Sun, Lei and Bogos, Bal{\'a}zs and Bonhoeffer, Sebastian},
  journal={Proceedings of the National Academy of Sciences of the U.S.A.},
  volume={118},
  number={13},
  pages={e2023467118},
  year={2021},
  publisher={National Academy of Sciences}
}

@article{miethke2021towards,
  title={Towards the sustainable discovery and development of new antibiotics},
  author={Miethke, Marcus and Pieroni, Marco and Weber, Tilmann and Br{\"o}nstrup, Mark and Hammann, Peter and Halby, Ludovic and Arimondo, Paola B and Glaser, Philippe and Aigle, Bertrand and Bode, Helge B and others},
  journal={Nature Reviews Chemistry},
  volume={5},
  number={10},
  pages={726--749},
  year={2021},
  publisher={Nature Publishing Group UK London}
}

@article{daruka2025eskape,
  title={ESKAPE pathogens rapidly develop resistance against antibiotics in development in vitro},
  author={Daruka, Lejla and Czikkely, M{\'a}rton Simon and Szili, Petra and Farkas, Zolt{\'a}n and Balogh, D{\'a}vid and Gr{\'e}zal, G{\'a}bor and Maharramov, Elvin and Vu, Thu-Hien and Sipos, Levente and Juh{\'a}sz, Szilvia and others},
  journal={Nature Microbiology},
  volume={10},
  number={2},
  pages={313--331},
  year={2025},
  publisher={Nature Publishing Group UK London}
}

@article{liu2023deep,
  title={Deep learning-guided discovery of an antibiotic targeting Acinetobacter baumannii},
  author={Liu, Gary and Catacutan, Denise B and Rathod, Khushi and Swanson, Kyle and Jin, Wengong and Mohammed, Jody C and Chiappino-Pepe, Anush and Syed, Saad A and Fragis, Meghan and Rachwalski, Kenneth and others},
  journal={Nature Chemical Biology},
  volume={19},
  number={11},
  pages={1342--1350},
  year={2023},
  publisher={Nature Publishing Group US New York}
}

@article{martins2025antibiotic,
  title={Antibiotic candidates for Gram-positive bacterial infections induce multidrug resistance},
  author={Martins, Ana and Jud{\'a}k, Fanni and Farkas, Zolt{\'a}n and Szili, Petra and Gr{\'e}zal, G{\'a}bor and Cs{\"o}rg{\H{o}}, B{\'a}lint and Czikkely, M{\'a}rton Simon and Maharramov, Elvin and Daruka, Lejla and Spohn, R{\'e}ka and others},
  journal={Science Translational Medicine},
  volume={17},
  number={780},
  pages={eadl2103},
  year={2025},
  publisher={American Association for the Advancement of Science}
}

@article{roemhild2018cellular,
  title={Cellular hysteresis as a principle to maximize the efficacy of antibiotic therapy},
  author={Roemhild, Roderich and Gokhale, Chaitanya S and Dirksen, Philipp and Blake, Christopher and Rosenstiel, Philip and Traulsen, Arne and Andersson, Dan I and Schulenburg, Hinrich},
  journal={Proceedings of the National Academy of Sciences of the U.S.A.},
  volume={115},
  number={39},
  pages={9767--9772},
  year={2018},
  publisher={National Academy of Sciences}
}

@article{sanz2023translating,
  title={Translating eco-evolutionary biology into therapy to tackle antibiotic resistance},
  author={Sanz-Garc{\'\i}a, Fernando and Gil-Gil, Teresa and Laborda, Pablo and Blanco, Paula and Ochoa-S{\'a}nchez, Luz-Edith and Baquero, Fernando and Mart{\'\i}nez, Jos{\'e} Luis and Hernando-Amado, Sara},
  journal={Nature Reviews Microbiology},
  volume={21},
  number={10},
  pages={671--685},
  year={2023},
  publisher={Nature Publishing Group UK London}
}

@article{beardmore2010rotating,
  title={Rotating antibiotics selects optimally against antibiotic resistance, in theory},
  author={Beardmore, Robert E and Pe{\~n}a-Miller, Rafael},
  journal={Mathematical Biosciences and Engineering},
  volume={7},
  number={3},
  pages={527--552},
  year={2010}
}

@article{beardmore2010antibiotic,
  title={Antibiotic cycling versus mixing: the difficulty of using mathematical models to definitively quantify their relative merits.},
  author={Beardmore, RE and Pena-Miller, R},
  journal={Mathematical Biosciences and Engineering},
  volume={7},
  number={4},
  pages={923--933},
  year={2010}
}

@article{nyhoegen2023sequential,
  title={Sequential antibiotic therapy in the laboratory and in the patient},
  author={Nyhoegen, Christin and Uecker, Hildegard},
  journal={Journal of the Royal Society Interface},
  volume={20},
  number={198},
  pages={20220793},
  year={2023},
  publisher={The Royal Society}
}

@article{muller2004issues,
  title={Issues in pharmacokinetics and pharmacodynamics of anti-infective agents: distribution in tissue},
  author={M{\"u}ller, Markus and dela Pena, Amparo and Derendorf, Hartmut},
  journal={Antimicrobial Agents and Chemotherapy},
  volume={48},
  number={5},
  pages={1441},
  year={2004}
}

@article{allegranzi2002impact,
  title={Impact of antibiotic changes in empirical therapy on antimicrobial resistance in intensive care unit-acquired infections},
  author={Girardini, F and Antozzi, L and Raiteri, R},
  journal={Journal of Hospital Infection},
  volume={52},
  number={2},
  pages={136--140},
  year={2002},
  publisher={Elsevier}
}

@article{nichol2015steering,
  title={Steering evolution with sequential therapy to prevent the emergence of bacterial antibiotic resistance},
  author={Nichol, Daniel and Jeavons, Peter and Fletcher, Alexander G and Bonomo, Robert A and Maini, Philip K and Paul, Jerome L and Gatenby, Robert A and Anderson, Alexander RA and Scott, Jacob G},
  journal={PLoS Computational Biology},
  volume={11},
  number={9},
  pages={e1004493},
  year={2015},
  publisher={Public Library of Science San Francisco, CA USA}
}

@article{sorum2022evolutionary,
  title={Evolutionary instability of collateral susceptibility networks in ciprofloxacin-resistant clinical Escherichia coli strains},
  author={S{\o}rum, Vidar and {\O}ynes, Emma L and M{\o}ller, Anna S and Harms, Klaus and Samuelsen, {\O}rjan and Podnecky, Nicole L and Johnsen, P{\aa}l J},
  journal={mBio},
  volume={13},
  number={4},
  pages={e00441--22},
  year={2022},
  publisher={American Society for Microbiology 1752 N St., NW, Washington, DC}
}

@article{hernando2023rapid,
  title={Rapid phenotypic convergence towards collateral sensitivity in clinical isolates of Pseudomonas aeruginosa presenting different genomic backgrounds},
  author={Hernando-Amado, Sara and L{\'o}pez-Causap{\'e}, Carla and Laborda, Pablo and Sanz-Garc{\'\i}a, Fernando and Oliver, Antonio and Mart{\'\i}nez, Jos{\'e} Luis},
  journal={Microbiology Spectrum},
  volume={11},
  number={1},
  pages={e02276--22},
  year={2023},
  publisher={American Society for Microbiology 1752 N St., NW, Washington, DC}
}

@article{hernando2025ciprofloxacin,
  title={Ciprofloxacin resistance rapidly declines in nfxB defective clinical strains of Pseudomonas aeruginosa},
  author={Hernando-Amado, Sara and Laborda, Pablo and La Rosa, Ruggero and Molin, S{\o}ren and Johansen, Helle Krogh and Mart{\'\i}nez, Jos{\'e} Luis},
  journal={Nature Communications},
  volume={16},
  number={1},
  pages={4992},
  year={2025},
  publisher={Nature Publishing Group UK London}
}

@article{morsky2022suppressing,
  title={Suppressing evolution of antibiotic resistance through environmental switching},
  author={Morsky, Bryce and Vural, Dervis Can},
  journal={Theoretical Ecology},
  volume={15},
  number={2},
  pages={115--127},
  year={2022},
  publisher={Springer}
}

@article{katriel2024optimizing,
  title={Optimizing antimicrobial treatment schedules: some fundamental analytical results},
  author={Katriel, Guy},
  journal={Bulletin of Mathematical Biology},
  volume={86},
  number={1},
  pages={1},
  year={2024},
  publisher={Springer}
}

@article{Czuppon2023,
  author       = {Czuppon, P. and Day, T. and Débarre, F. and Blanquart, F.},
  title        = {A stochastic analysis of the interplay between antibiotic dose, mode of action, and bacterial competition in the evolution of antibiotic resistance},
  journal      = {PLoS Computational Biology},
  volume       = {19},
  number       = {8},
  pages        = {e1011364},
  year         = {2023}
}

@article{yoshida2017time,
  title={Time-programmable drug dosing allows the manipulation, suppression and reversal of antibiotic drug resistance in vitro},
  author={Yoshida, Mari and Reyes, Sabrina Galinanes and Tsuda, Soichiro and Horinouchi, Takaaki and Furusawa, Chikara and Cronin, Leroy},
  journal={Nature Communications},
  volume={8},
  number={1},
  pages={15589},
  year={2017},
  publisher={Nature Publishing Group UK London}
}

@article{barbosa2019evolutionary,
  title={Evolutionary stability of collateral sensitivity to antibiotics in the model pathogen Pseudomonas aeruginosa},
  author={Barbosa, Camilo and Roemhild, Roderich and Rosenstiel, Philip and Schulenburg, Hinrich},
  journal={Elife},
  volume={8},
  pages={e51481},
  year={2019},
  publisher={eLife Sciences Publications, Ltd}
}

@article{GilGil2024SCV_HR,
  author  = {Gil-Gil, Teresa and Berryhill, Brandon A. and Manuel, Joshua A. and Smith, Andrew P. and McCall, Ingrid C. and Baquero, Fernando and Levin, Bruce R.},
  title   = {The evolution of heteroresistance via small colony variants in \textit{Escherichia coli} following long-term exposure to bacteriostatic antibiotics},
  journal = {Nature Communications},
  year    = {2024},
  volume  = {15},
  pages   = {7936}
}

@article{Ankomah2014Collaboration,
  author  = {Ankomah, Peter and Levin, Bruce R.},
  title   = {Exploring the collaboration between antibiotics and the immune response in the treatment of acute, self-limiting infections},
  journal = {Proceedings of the National Academy of Sciences of the U.S.A.},
  year    = {2014},
  volume  = {111},
  number  = {21},
  pages   = {8331--8338}
}

@article{molina-hernandez2026github,
  author       = {Molina-Hernández, Javier},
  title        = {Optimization of sequential therapies to maximize extinction of resistant bacteria through collateral sensitivity},
  year         = {2026},
  journal    = {Zenodo},
  doi          = {10.5281/zenodo.18656444},
  url          = {https://doi.org/10.5281/zenodo.18656444},
}

@article{shepherd2024ecological,
  title={Ecological and evolutionary mechanisms driving within-patient emergence of antimicrobial resistance},
  author={Shepherd, Matthew J and Fu, Taoran and Harrington, Niamh E and Kottara, Anastasia and Cagney, Kendall and Chalmers, James D and Paterson, Steve and Fothergill, Joanne L and Brockhurst, Michael A},
  journal={Nature Reviews Microbiology},
  volume={22},
  number={10},
  pages={650--665},
  year={2024},
  publisher={Nature Publishing Group UK London}
}

@article{bonhoeffer1997evaluating,
  title={Evaluating treatment protocols to prevent antibiotic resistance},
  author={Bonhoeffer, Sebastian and Lipsitch, Marc and Levin, Bruce R},
  journal={Proceedings of the National Academy of Sciences of the U.S.A.},
  volume={94},
  number={22},
  pages={12106--12111},
  year={1997},
  publisher={The National Academy of Sciences of the USA}
}

@article{muetter2024impact,
  title={The impact of treatment strategies on the epidemiological dynamics of plasmid-conferred antibiotic resistance},
  author={Muetter, Malte and Angst, Daniel C and Regoes, Roland R and Bonhoeffer, Sebastian},
  journal={Proceedings of the National Academy of Sciences of the U.S.A.},
  volume={121},
  number={52},
  pages={e2406818121},
  year={2024},
  publisher={National Academy of Sciences}
}

@article{blair2015acrb,
  title={AcrB drug-binding pocket substitution confers clinically relevant resistance and altered substrate specificity},
  author={Blair, Jessica MA and Bavro, Vassiliy N and Ricci, Vito and Modi, Niraj and Cacciotto, Pierpaolo and Kleinekath\"ofer, Ulrich and Ruggerone, Paolo and Vargiu, Attilio V and Baylay, Alison J and Smith, Helen E and others},
  journal={Proceedings of the National Academy of Sciences of the U.S.A.},
  volume={112},
  number={11},
  pages={3511--3516},
  year={2015},
  publisher={National Academy of Sciences}
}

@article{mwangi2007tracking,
  title={Tracking the in vivo evolution of multidrug resistance in Staphylococcus aureus by whole-genome sequencing},
  author={Mwangi, Michael M and Wu, Shang Wei and Zhou, Yanjiao and Sieradzki, Krzysztof and de Lencastre, Herminia and Richardson, Paul and Bruce, David and Rubin, Edward and Myers, Eugene and Siggia, Eric D and others},
  journal={Proceedings of the National Academy of Sciences of the U.S.A.},
  volume={104},
  number={22},
  pages={9451--9456},
  year={2007},
  publisher={National Academy of Sciences}
}

@article{wheatley2021rapid,
  title={Rapid evolution and host immunity drive the rise and fall of carbapenem resistance during an acute Pseudomonas aeruginosa infection},
  author={Wheatley, Rachel and Diaz Caballero, Julio and Kapel, Natalia and De Winter, Fien HR and Jangir, Pramod and Quinn, Angus and del Barrio-Tofino, Ester and Lopez-Causape, Carla and Hedge, Jessica and Torrens, Gabriel and others},
  journal={Nature Communications},
  volume={12},
  number={1},
  pages={2460},
  year={2021},
  publisher={Nature Publishing Group UK London}
}

@article{chung2022rapid,
  title={Rapid expansion and extinction of antibiotic resistance mutations during treatment of acute bacterial respiratory infections},
  author={Chung, Hattie and Merakou, Christina and Schaefers, Matthew M and Flett, Kelly B and Martini, Sarah and Lu, Roger and Blumenthal, Jennifer A and Webster, Shanice S and Cross, Ashley R and Al Ahmar, Roy and others},
  journal={Nature Communications},
  volume={13},
  number={1},
  pages={1231},
  year={2022},
  publisher={Nature Publishing Group UK London}
}

@article{Pluchino2012,
  author    = {Pluchino, Kristen M. and Hall, Matthew D. and Goldsborough, Andrew S. and Callaghan, Richard and Gottesman, Michael M.},
  title     = {Collateral sensitivity as a strategy against cancer multidrug resistance},
  journal   = {Drug Resistance Updates},
  year      = {2012},
  volume    = {15},
  number    = {1--2},
  pages     = {98--105}
}

@article{Dhawan2017,
  author    = {Dhawan, Andrew and Nichol, Daniel and Kinose, Fumi and Abazeed, Mohamed E and Marusyk, Andriy and Haura, Eric B and Scott, Jacob G},
  title     = {Collateral sensitivity networks reveal evolutionary instability and novel treatment strategies in {ALK} mutated non-small cell lung cancer},
  journal   = {Scientific Reports},
  year      = {2017},
  volume    = {7},
  pages     = {1232}
}

@article{Zhao2016,
  author    = {Zhao, Boyang and Sedlak, Joseph C. and Srinivas, Raja and Creixell, Pau and Pritchard, Justin R. and Tidor, Bruce and Lauffenburger, Douglas A. and Hemann, Michael T.},
  title     = {Exploiting temporal collateral sensitivity in tumor clonal
               evolution},
  journal   = {Cell},
  year      = {2016},
  volume    = {165},
  number    = {1},
  pages     = {234--246}
}

@article{Gatenby2009,
  author    = {Gatenby, Robert A. and Silva, Ariosto S. and Gillies, Robert J. and Frieden, B. Roy},
  title     = {Adaptive therapy},
  journal   = {Cancer Research},
  year      = {2009},
  volume    = {69},
  number    = {11},
  pages     = {4894--4903}
}

@article{Loria2022,
  author    = {Loria, Rossella and Vici, Patrizia and {Di Lisa}, Francesca Sofia and Soddu, Silvia and {Maugeri-Sacc\`{a}}, Marcello and Bon, Giulia},
  title     = {Cross-resistance among sequential cancer therapeutics: an emerging issue},
  journal   = {Frontiers in Oncology},
  year      = {2022},
  volume    = {12},
  pages     = {877380}
}

@article{Danisik2023,
  author    = {Danisik, Nurseda and {Celikbas Yilmaz}, Kubra and Acar, Ahmet},
  title     = {Identification of collateral sensitivity and evolutionary landscape of chemotherapy-induced drug resistance using cellular barcoding technology},
  journal   = {Frontiers in Pharmacology},
  year      = {2023},
  volume    = {14},
  pages     = {1178489}
}

@article{West2020,
  author    = {West, Jeffrey and You, Li and Zhang, Jingsong and Gatenby, Robert A. and Brown, Joel S. and Newton, Paul K. and Anderson, Alexander R. A.},
  title     = {Towards multidrug adaptive therapy},
  journal   = {Cancer Research},
  year      = {2020},
  volume    = {80},
  number    = {7},
  pages     = {1578--1589}
}

@article{Musken2018-tz,
  title     = "Breaking the vicious cycle of antibiotic killing and regrowth of
               biofilm-residing Pseudomonas aeruginosa",
  author    = "M{\"u}sken, Mathias and Pawar, Vinay and Schwebs, Timo and
               B{\"a}hre, Heike and Felgner, Sebastian and Weiss, Siegfried and
               H{\"a}ussler, Susanne",
  journal   = "Antimicrob. Agents Chemother.",
  publisher = "American Society for Microbiology",
  volume    =  62,
  number    =  12,
  month     =  dec,
  year      =  2018
}

@article{rojo2016sequential,
  title={Sequential treatment of biofilms with aztreonam and tobramycin is a novel strategy for combating Pseudomonas aeruginosa chronic respiratory infections},
  author={Rojo-Molinero, Estrella and Maci{\`a}, Mar{\'\i}a D and Rubio, Rosa and Moy{\`a}, Bartolom{\'e} and Cabot, Gabriel and L{\'o}pez-Causap{\'e}, Carla and P{\'e}rez, Jos{\'e} L and Cant{\'o}n, Rafael and Oliver, Antonio},
  journal={Antimicrobial Agents and Chemotherapy},
  volume={60},
  number={5},
  pages={2912--2922},
  year={2016},
  publisher={American Society for Microbiology 1752 N St., NW, Washington, DC}
}

\end{document}